\documentclass[prd,superscriptaddress,amsmath,twocolumn,nofootinbib]{revtex4-2}

\pdfoutput=1
\usepackage{times}

\usepackage{graphicx}
\usepackage{xcolor}
\usepackage{braket}
\usepackage{array}
\usepackage{booktabs}
\usepackage{enumitem}
\usepackage{epsf,latexsym,bbm,euscript}
\usepackage{cancel}
\usepackage{mathrsfs,mathtools,amsmath,amssymb}
\makeatletter
\newcommand*{\defeq}{\mathrel{\rlap{%
			\raisebox{0.3ex}{$\m@th\cdot$}}%
		\raisebox{-0.3ex}{$\m@th\cdot$}}%
	=}
\newcommand*{\eqdef}{=\mathrel{\rlap{%
			\raisebox{0.3ex}{$\m@th\cdot$}}%
		\raisebox{-0.3ex}{$\m@th\cdot$}}%
}
\makeatother

\newcommand{\RNum}[1]{\uppercase\expandafter{\romannumeral #1\relax}}

\usepackage{scalerel}
\usepackage{tikz}
\usetikzlibrary{svg.path}
\definecolor{orcidlogocol}{HTML}{A6CE39}
\tikzset{
	orcidlogo/.pic={
		\fill[orcidlogocol] svg{M256,128c0,70.7-57.3,128-128,128C57.3,256,0,198.7,0,128C0,57.3,57.3,0,128,0C198.7,0,256,57.3,256,128z};
		\fill[white] svg{M86.3,186.2H70.9V79.1h15.4v48.4V186.2z}
		svg{M108.9,79.1h41.6c39.6,0,57,28.3,57,53.6c0,27.5-21.5,53.6-56.8,53.6h-41.8V79.1z M124.3,172.4h24.5c34.9,0,42.9-26.5,42.9-39.7c0-21.5-13.7-39.7-43.7-39.7h-23.7V172.4z}
		svg{M88.7,56.8c0,5.5-4.5,10.1-10.1,10.1c-5.6,0-10.1-4.6-10.1-10.1c0-5.6,4.5-10.1,10.1-10.1C84.2,46.7,88.7,51.3,88.7,56.8z};
	}
}
\newcommand\orcidlink[1]{\href{https://orcid.org/#1}{\mbox{\scalerel*{
				\begin{tikzpicture}[yscale=-1,transform shape]
					\pic{orcidlogo};
				\end{tikzpicture}
			}{X}}}}
\newcommand{\be}{\begin{equation}}
\newcommand{\ee}{\end{equation}}

\usepackage{soul}

\usepackage{url,hyperref}
\hypersetup{colorlinks,linkcolor={blue!55!black},citecolor={red!50!black},urlcolor={blue!45!black},breaklinks=true}

\begin{document}
	
	\title{Euclidean and Hamiltonian thermodynamics for regular black holes}

		\author{Fil Simovic\orcidlink{0000-0003-1736-8779}}
	\email{fil.simovic@mq.edu.au}
	\affiliation{School of Mathematical and Physical Sciences, Macquarie University, Sydney, New South Wales 2109, Australia}

	\author{Ioannis Soranidis\orcidlink{0000-0002-8652-9874}}
	\email{ioannis.soranidis@hdr.mq.edu.au}
	\affiliation{School of Mathematical and Physical Sciences, Macquarie University, Sydney, New South Wales 2109, Australia}

	\begin{abstract}
	
	We investigate the thermodynamic properties of the Hayward regular black hole using both Euclidean path integral and Hamiltonian methods, in asymptotically anti-de Sitter, Minkowski, and de Sitter spacetimes. With the inclusion of matter fields which act as a source for the regular black hole geometry, an effective temperature emerges that differs from the conventional definition related to the Killing surface gravity. We posit that this temperature is the appropriate choice for studying thermodynamic phenomena, by demonstrating consistency between the Euclidean and Hamiltonian formulations in the appropriate limits.  We examine the thermodynamic properties and phase structure of the Hayward black hole in the canonical ensemble and show that, counter to some earlier indications, standard mean-field theory critical behavior is observed when the cosmological constant is treated as a thermodynamic pressure. We note the absence of a Hawking-Page transition, and conjecture that quantum gravity corrections which are suitably strong to regulate the Schwarzschild singularity generically prevent the transition from occurring. We also show that the Smarr relation remains linear in all cases, despite the absence of a linearity proof for nonlinear electrodynamic theories with nonsymmetry inheriting fields.

	\end{abstract}
	
	\maketitle
	
	\section{Introduction}

    The singular nature of classical black hole solutions in general relativity remains one of the most important yet obtrusive features of prototypical black hole models. Signaling a breakdown of the effective field theory description of quantum gravity near the Planck scale, such singularities are believed to be largely provisional. Yet, how exactly a tentative quantum theory of gravity regulates the singularity, whether such a smoothing procedure propagates its effects beyond the Planck scale, and whether indeed the singularity can be tamed without invoking sub-Planckian or quantum degrees of freedom, remain important open questions.
    \\
    
    The smoothing of the central singularity is inextricably tied to energy conditions. The original Hawking-Penrose singularity theorems achieve the required geodesic focusing via the strong energy condition (SEC), so naturally early regular black hole models such as the Bardeen black hole precisely drop this assumption on the classical collapsing matter. Since it is known that both classical and quantum field theory can violate the strong energy condition, the singularity theorems have since been revisited in a number of forms using the averaged null energy condition, weak energy condition, and various quantum energy inequalities in place of the SEC. A host of regular black hole models that generally involve violation of one or more of these energy conditions have come about since the introduction of the Bardeen black hole \cite{B:68}, including phantom black holes \cite{bronnikov2006}, noncommutative black holes \cite{nicolini2006} and the Hayward model \cite{H:06}. 
    \\
    
    Among the available singularity-free black hole models, certain candidates distinguish themselves by being generated by classical matter distributions coupled minimally to Einstein-Hilbert gravity. In particular, the Bardeen and Hayward metrics (as well as various extensions) can be sourced by electric and/or magnetic charges in general relativity coupled to nonlinear electrodynamics \cite{ayon-beato2000,fan2017}. The nonlinear electrodynamic (NED) Lagrangians involved remain important sources in the context of string theory and the study of various condensed matter systems (see \cite{sorokin2022} for a recent review). 
    \\
    
    The nature of how regular black holes (RBHs) are sourced at the classical level has led to widespread disagreement on how to properly formulate the laws of thermodynamics for RBHs. Whenever new parameters enter into the Lagrangian, extra terms ostensibly appear in the mechanical first law which do not always lend themselves to an obvious thermodynamic interpretation. Additionally, some approaches maintain that the entropy is the Bekenstein-Hawking result of $S=A/4$, while others argue that corrections to the entropy arise \cite{N:18}. The issue becomes even more subtle in the case of the Hayward model, since properly incorporating matter fields (which are required to generate the geometry) in the first law is highly nontrivial. Understanding the thermodynamic properties of regular black hole solutions is also important for studying how regularization of the central singularity propagates into the classical sector, since these solutions can be viewed as classical approximations to a regularized geometry arising from quantum gravity theory. Especially in the asymptotically anti-de Sitter case, we expect important modifications to the dual gauge theory description in the context of the anti-de Sitter/conformal field theory correspondence (AdS/CFT) \cite{maldacena1999}.
    \\
    
    There are a variety of methods available for studying black hole thermodynamics which have been developed since the pioneering work of Bekenstein, Hawking, and collaborators \cite{BCH:73}. Simply computing the entropy of a black hole may be done using the covariant phase space approach of Iyer and Wald \cite{IW:94}, the Euclidean path integral of Gibbons and Hawking \cite{gibbons1977}, the conical deficit formula of Susskind \cite{susskind1994}, various entanglement entropy computations \cite{solodukhin2011}, CFT techniques \cite{carlip1999}, and more. While all of these seem to reproduce the famous $S=A/4$ result for black holes in Einstein-Hilbert gravity, not all of these techniques and their associated first law constructions are equivalent, and not all can be applied in the same contexts. For the Hayward regular black hole, which is sourced by a gauge field with nonlinear Lagrangian, it is especially unclear which method should be preferred. Over the years, a number of investigations have been carried out using different methods. The Hayward-AdS model was studied previously in \cite{fan2017} and \cite{luo2023} using variables defined through a Hamiltonian variation of the Komar mass, though in the former case we believe an unsuitable choice of thermodynamic variables has led to an unexpected departure from the expected behavior of AdS black holes in the extended phase space. The asymptotically flat case was considered in \cite{molina2021}, taking $S=A/4$ as a given and assuming a Gibbsian thermodynamic interpretation from the onset, though no mechanism was given for establishing thermodynamic equilibrium, so it is not clear what ensemble is being defined there. The case of general $\Lambda$ was considered in \cite{LYGM:23} using a Euclidean approach, but only the action was computed and no further thermodynamic analysis was done. 
    \\
    
    In this paper we aim to shed light on this state of affairs and resolve some of the disagreement between various implementations of the thermodynamical laws for regular black holes. We consider the Hayward black hole model in asymptotically flat, anti-de Sitter, and de Sitter spacetimes, and attempt to demonstrate consistency between the Hamiltonian and Euclidean path integral formulations where direct comparisons between the methods are possible. Properly accounting for the variation of $\Lambda$, the correct definition of mass for the Hayward black hole, and accounting for the mechanisms required to establish equilibrium for different asymptotic behavior, we examine the phase structure and thermodynamic stability of the various solutions. In the AdS case, our work differs from that of \cite{fan2017} and \cite{luo2023} in the choice of thermodynamic parameters, which in the former we believe have been incorrectly identified. As a result, we are able to show that the Hayward-AdS black hole does indeed possess the expected mean-field theory critical exponents.
    \\
    This paper is organized as follows: in Sec.~\ref{sec:2} we describe the general structure of NED theories, and demonstrate how the Hayward model arises, including its properties, source, and de Sitter embedding. In Sec.~\ref{thermo}, we discuss how the first law of black hole thermodynamics and Smarr relation apply to regular black holes and demonstrate consistency between Hamiltonian and Euclidean path integral frameworks. In Sec.~\ref{sec:phase-structure} we analyze the phase structure of the Hayward model. Specifically, in Sec.~\ref{ads} we study the Hayward-AdS black hole, examining its thermodynamic stability, phase structure, and behavior near the critical point and in Secs.~\ref{flat} and \ref{desitter} we consider the asymptotically Minkowski and de Sitter cases respectively. We conclude in Sec.~\ref{discussion} with a discussion of the main results and implications for future investigations. Throughout, we work in $d=4$ and units where $\hbar=c=G=1$ are used unless otherwise noted.
	
	\section{Nonlinear Electrodynamics}\label{sec:2}
	
	 Regular black hole geometries generically require additional matter fields to source their non-singular metrics, with the first example of an exact solution being given by \cite{ABG:98} using a nonlinear electrodynamic (NED) source. That NED can generate other well-known regular geometries was later demonstrated for the Bardeen black hole in \cite{ayon-beato2000} and further generalized to a variety of two-parameter families of spherically symmetric RBHs solutions in \cite{FW:16}. The geometry and dynamics are determined by the action of the given NED theory, which in the presence of a cosmological constant $\Lambda$ has the general form
	 \begin{align}
	 	I=\frac{1}{16\pi}\int d^4x\sqrt{-g}\left[R-2\Lambda+4\mathcal{L}(\mathcal{F},\mathcal{G})\right],\label{eq:action}
	 \end{align}
	 where $g=\det{g_{\mu\nu}}$, $R$ is the Ricci scalar, and $\mathcal{L}(\mathcal{F},\mathcal{G})$ is the Lagrangian density of the NED theory under consideration. $\mathcal{L}$ is a function of two electromagnetic invariants $\mathcal{F}\equiv F^{\mu\nu}F_{\mu\nu}$ and $\mathcal{G}\equiv F_{\mu\nu} \star\! F^{\mu\nu}$, which are functions of the field strength tensor $F_{\mu\nu}=\partial_{\mu}A_{\nu}-\partial_{\nu}A_{\mu}$. It may also be a function of a finite number of additional real parameters $\{\beta_i\}$. An auxiliary two-form can be defined as
	 \be
	 Z_{\mu \nu} \equiv-4\left(\dfrac{\partial \mathcal{L}}{\partial \mathcal{F}}F_{\mu \nu}+\dfrac{\partial \mathcal{L}}{\partial \mathcal{G}}\star\!F_{\mu \nu}\right),
	 \ee
	 so that the equations of motion of the theory are the generalized source-free Einstein-Maxwell equations
	 \be
	 G_{\mu\nu}=8\pi T_{\mu\nu}\ ,\quad dF=0\ ,\quad d\star\!Z =0\ .
	 \ee
	 The Maxwell case corresponds to $\mathcal{L}(\mathcal{F},\mathcal{G})=\mathcal{L}(\mathcal{F})=-\mathcal{F}/4$.
	 In general, the NED theory defined by \eqref{eq:action} can possess both electric and magnetic charges, with the existence of an electromagnetic duality $F\rightarrow \star F$ being tied to the existence of a Maxwell limit. The electric and magnetic charges are defined, respectively, as Komar integrals over a smooth closed 2-surface $\mathcal{S}$ as 
	 \be
	 Q_e\equiv\dfrac{1}{4\pi}\int_\mathcal{S}\star Z\ ,\quad Q_m\equiv\dfrac{1}{4\pi}\int_\mathcal{S}F\ , 
	 \ee
	 although many models can be generated by either electric or magnetic charge alone. For Lagrangians which have a Maxwellian weak-field limit, solutions {\it must} be sourced entirely by magnetic charge, with electrically sourced solutions generically requiring a different form of Lagrangian in the near and far regions \cite{bronnikov2017}. A subclass of NED theories defined by \eqref{eq:action} possess invariance under an electromagnetic duality transformation by
	 \be
	 \begin{aligned}
	 F &\rightarrow F \cos \alpha + \star Z \sin \alpha, \\
	 Z &\rightarrow Z \cos \alpha + \star F \sin \alpha \ 
	 ,
	 \end{aligned}
 \ee
 examples of which are Born-Infeld \cite{BI:34}, ModMax \cite{bandos2020}, and power-Maxwell theories \cite{hassaine2007} along with standard Maxwell electrodynamics. The existence of such a duality is important for establishing linearity of the Smarr relation for such theories \cite{bokulic2021}, while theories without the duality may exhibit nonlinear Smarr relations instead. These NED theories remain of broad theoretical interest for many reasons, having been introduced to regularize the classical divergences associated with point charges, and also arising as high-energy corrections to standard electrodynamics from more fundamental theories.

	 \subsection{The Hayward model}
	 One particularly important example of a geometry generated by NED theory is the Hayward regular black hole \cite{H:06}. Regular black holes are of themselves of significant theoretical interest, representing classical black hole geometries with no central singularity. This is a highly non-trivial condition that entails finiteness of all algebraic curvature invariants at the center as well as flatness. A necessary requirement for regularity at the center is the existence of a de Sitter or anti-de Sitter core. The usual $r=0$ singularity common to textbook black hole solutions to general relativity signals a breakdown of the effective field theory description of general relativity and is expected to be resolved by an appropriate quantum theory of gravity. Classical regular black hole solutions therefore represent a first-order approximation to the geometry that results from whatever regularization procedure inevitably smooths the central singularity. It is therefore of significant theoretical interest to study the properties of regular black hole solutions. The Hayward model represents one such solution, and can be generated from the following Lagrangian density
	 \begin{align}
	 	\mathcal{L}(\mathcal{F})=\frac{12}{\alpha}\frac{(\alpha \mathcal{F})^{3/2}}{(1+(\alpha \mathcal{F})^{3/4})^2}\ , \label{eq:LF}
	 \end{align} 
	 where $\alpha$ is one of the free parameters of the theory \cite{FW:16}. This theory is independent of $\mathcal{G}$ and  possesses only magnetic charge. The vector potential is given explicitly by 
	 \begin{align}
	 	A_{\mu}=\left(0,0,0,Q_{m}\cos{\theta}\right)\ ,\label{eq:Am}
	 \end{align} 
	  and the variation of the action \eqref{eq:action} leads to 
	  \begin{align}
	  	G_{\mu\nu}=8\pi T_{\mu\nu}\quad  \text{and} \quad \nabla_{\mu}\left(\frac{\partial \mathcal{L}}{\partial \mathcal{F}}F^{\mu\nu}\right)=0\ ,
	  \end{align}
	 where $G_{\mu\nu}$ is the Einstein tensor in the presence of a cosmological constant and the source term $T_{\mu\nu}$ is given by 
	 \begin{align}
	 	T_{\mu\nu}=\frac{1}{4\pi}\left(\frac{\partial \mathcal{L}}{\partial \mathcal{F}} F_{\mu\lambda}F^{\lambda}_{\nu}-\frac{1}{4}g_{\mu\nu}\mathcal{L}\right).
	 \end{align}
	 The equations of motion admit a spherically symmetric geometry
	 	\begin{align}
	 	ds^2=-f(r)dt^2+\frac{dr^2}{f(r)}+r^2d\Omega_{2}\ ,
	 \end{align}
	 where $d\Omega_{2}$ is the metric on $\mathcal{S}^2$ and the metric function $f(r)$ is given by 
	 \begin{align}
	 	f(r)=1-\frac{2\alpha^{-1}q^3r^2}{r^3+q^3}-\frac{\Lambda}{3}r^2\ , \label{eq:faq}
	 \end{align}
	 where $\Lambda$ is the cosmological constant, and $q$ is an integration constant related to the magnetic charge $Q_{m}$ and coupling $\alpha$ of the theory through
	 \begin{align}
	 	Q_{m}=\frac{q^2}{\sqrt{2\alpha}}\ .\label{eq:Qm}
	 \end{align}
	 In the notation of \cite{H:06}, the metric function of the regular spherically symmetric geometry is given by
	   \begin{align}
	 	f(r)=1-\frac{2mr^2}{r^3+2ml^2}-\frac{\Lambda}{3}r^2\ ,\label{eq:f-Hayward}
	 \end{align}
	 where $m$ is the mass parameter and $l$ is a minimal length scale characterizing the size of the regular center. When $l\rightarrow 0$ this reduces to the Schwarzschild-anti-de Sitter (or -de Sitter) metric. Direct comparison to \eqref{eq:faq} allows one to identify $q$ and $\alpha$ appearing in the NED metric as \cite{F:16,SZ:22} 
	 \begin{align}
	 	\alpha=2l^2\quad \text{and}\quad q^3=2ml^2\ . \label{eq:alpha-q}
	 \end{align}
	Therefore the regularization of the central singularity, which is characterized by a minimal length scale $l$, can be given an effective description in terms of a nonlinear electromagnetic field sourced by magnetic charge. 
	\\
	
	A few comments are warranted concerning the Hayward model \eqref{eq:f-Hayward}. First, the solution is sourced entirely by magnetic charge, which to date has not been observed in the Universe despite a number of significant observational efforts  \cite{themacrocollaboration2002,balestra2008}. Nonetheless, heavy magnetic monopoles which form in the early Universe remain a generic prediction of Grand Unified Theories \cite{hooft1974,polyakov1974,kibble1980}, where they arise from spontaneous symmetry breaking at the GUT scale $\sim 10^{16}$ GeV, and have important implications for structure formation and early Universe cosmology (see \cite{preskill1984} for a review). Second, one may readily observe that in the weak-field limit the Lagrangian \eqref{eq:LF} does not approach the Maxwell limit but rather $\mathcal{L}\sim \sqrt{\alpha} \mathcal{F}^{3/2}$ which is stronger than the Maxwell field. While this feature is somewhat unattractive, the Hayward model nonetheless represents one of the very few available classical regular black hole geometries that one can use to model quantum gravity effects on the central singularity, since there are strict no-go theorems which forbid regular geometries in ordinary Einstein-Maxwell theory. Furthermore, nonlinear theories with pure electric source that have a Maxwell weak-field limit are also known to not admit solutions with regular center \cite{B:01}. The magnetically sourced Hayward model therefore serves as an extremely useful and minimal prototype model of a regular black hole.
	\\
	
	In what follows we use the metric in the form of \eqref{eq:f-Hayward}, considering the minimal length $l$ as a more fundamental parameter than $q$ and $\alpha$, which arise in just one of a distinct number of ways to generate the same geometry. This is in contrast to earlier work on the Hayward-AdS black hole \cite{fan2017}, whose results we regard with caution. Their work has treated $Q_m$ and $\alpha$ ($\sigma$ in their notation) as two independent parameters, when there is in fact only one (which we recast in terms of $l$) since $Q_m$ depends only on $\alpha$ and the ADM mass $m$. The previous work does not identify $m$ as the ADM mass, though it must be the case as pointed out initially by \cite{toshmatov2018}.

	\section{First law and Smarr formula} \label{thermo}
	\subsection{Hamiltonian methods}
	The first law of black hole mechanics occupies a significant portion of the theoretical physics landscape. First derived in \cite{BCH:73} as a relation between variations of the physical parameters describing a black hole in Einstein gravity, an analogous relation has since been shown to hold in any theory of gravity arising from a diffeomorphism-invariant Lagrangian \cite{W:93,IW:94}. The generalization of the first law to nonlinear electrodynamics was first studied in \cite{R:03}, although the Smarr formula \cite{S:73} was not found to be satisfied. This was later addressed in \cite{ZG:18} by appropriately accounting for the extra parameters appearing in the NED theory. The result is a relation resembling the ordinary first law with additional terms:
	\begin{align}
		dM=TdS+\Phi dQ_{e}+\Psi_{H}dQ_{m}+\sum_{i}K_{i}d\beta_{i}.\label{eq:1st law-gen}
	\end{align}
	In the above, $M$ is the Komar mass \cite{K:63} which coincides with the ADM mass \cite{ADM:59} in asymptotically flat spacetimes, $T$ is the Hawking temperature (as determined the surface gravity $\kappa$ through $T=\kappa/2\pi$), $S=A/4$ is the entropy, and $\{\Phi,\Psi_{H}\}$ are the electric and magnetic potentials associated with the electric and magnetic charges $\{Q_{e},Q_{m}\}$. The terms $K_{i}$ represent the potentials conjugate to the parameters $\beta_{i}$ of the theory. In \cite{ZG:18}, these quantities are explicitly computed for the Bardeen black hole and black hole solutions in Born-Infeld theory. The Smarr relation was also shown to hold for both cases. 
	\\
	
	Using the definitions of \cite{ZG:18}, we explicitly compute the thermodynamic quantities entering the first law for the Hayward black hole, both in the presence and absence of a cosmological constant $\Lambda$. The cosmological constant, when treated as an independently varying parameter, naturally enters into the first law as a work term $V dP$ with the identification $P=-\Lambda/8\pi$. In this extended phase space, the thermodynamic volume $V$ appears naturally conjugate to variations in $P$ \cite{kastor2009} and the mass $M$ is identified as a thermodynamic enthalpy \cite{KMT:17}. The variation of $P$, and therefore $\Lambda$, arises naturally in a consistent variational principle \cite{urano2009}, and can acquire a dynamical status through a 3-form gauge potential as in the Brown-Teitelboim mechanism \cite{BT:87}. The minimal length $l$ is also treated as a variable parameter, on the basis that it emerges in more fundamental theories from vacuum expectation values of the elementary fields \cite{GKK:96,CM:95}, and when back-reaction of the evaporation process is accounted for \cite{BBCRG:21,BBCRG:22,CRFLV:23}. The variation of these additional parameters is also required to establish the correct Smarr relation \cite{kastor2009,B:05}, which is widely believed to be universal and arises from quite general scaling arguments.
	\\
	
	We proceed with the explicit calculation of the quantities of the first law in the case of Hayward black hole embedded in a spacetime with a cosmological constant. In our case there is only a single additional parameter $\{\beta_{i}\}=\alpha$ appearing in \eqref{eq:1st law-gen}, along with its conjugate potential $K_{\alpha}$.  The first law therefore becomes
	\begin{align}
		dM=TdS+\Psi_{H}dQ_{m}+K_{\alpha}d\alpha+VdP\label{eq:1st law-gen-hayward}\ .
	\end{align}
	The mass parameter can be written in terms of the event horizon radius $r_{h}$ as 
	\begin{align}
		m=\frac{3r^3_h-\Lambda r^5_h}{6 r^2_h-2l^2(3-\Lambda r^2_h)}\ ,\label{eq:mass}
	\end{align}
    so that \eqref{eq:f-Hayward} can be expressed as
    \be
    f(r)=1-\frac{r^2 r_h^3 \left(3-\Lambda r_h^2\right)}{3 r^3 r_h^2-l^2 \left(3-\Lambda r_h^2\right) \left(r^3-r_h^3\right)}\ ,
    \ee
    where $r_h$ is the outermost real root of $f(r)=0$ (unless $\Lambda>0$). The Hawking temperature $T$ is determined by the Killing surface gravity $\kappa$ as
	\begin{align}
		T=\frac{\kappa}{2\pi}=\dfrac{f'(r)}{8\pi}\Big|_{r=r_h}=\frac{3r^2_h(1-r^2_h\Lambda)-l^2(3-r^2_h\Lambda)^2}{12\pi r^3_h}\label{eq:temp-surf.grav}\ ,
	\end{align}
    and the entropy is simply $S=A/4=\pi r_h^2$ \cite{B:72,B:73,B:74}. Since the spacetime is static, it admits a Killing vector $\xi^{\mu}=\partial_{t}$ through which the magnetic field $H_{\mu}$ is defined
    \begin{align}
    	H_{\mu}=-\star B_{\mu\nu}\xi^{\nu}\ ,\label{eq:H}
    \end{align}
    where
    \begin{align}
    	\star B_{\mu\nu}=\frac{1}{2}\epsilon_{\mu\nu\rho\sigma}B^{\rho\sigma}\quad \text{and}\quad B^{\mu\nu}=\mathcal{L}'(\mathcal{F})F^{\mu\nu}\ ,\label{eq:B:def}
    \end{align}
    with $\epsilon_{\mu\nu\rho\sigma}$ being the volume form. The magnetic potential $\Phi$ is related to the magnetic field through 
    \begin{align}
    	H_{\mu}=\nabla_{\mu}\Psi\ ,
    \end{align}
    whose integration along with the boundary condition $\lim_{\ r\rightarrow \infty}\Psi_{H}=0$ gives
    \begin{align}
    	\Psi(r)=\frac{3q^4(2r^3+q^3)}{\sqrt{2\alpha}(r^3+q^3)^2}\ .
    \end{align}
    The quantity $\Psi_H$ appearing in the first law is then simply $\Psi (r)$ evaluated on the horizon $r=r_h$. Finally, the conjugate potential $K_{\alpha}$ associated with the parameter $\alpha$ is given by \cite{R:03,ZG:18} 
    \begin{align}
    	K_{\alpha}=\frac{1}{4}\int_{r_h}^{\infty}\!r^2\ \frac{\partial \mathcal{L}}{\partial \alpha}\ ,\label{eq:Ka-int}
    \end{align}
    which upon integration gives
    \begin{align}
    	K_{\alpha}=\frac{6q^6(-2q^3+r^3_h)}{4\alpha^2(r^3_h+q^3)^2}\ .\label{eq:Ka}
    \end{align}
Finally, the geometric volume is $V=4\pi r_h^3/3$. It can be shown \cite{F:17} that these quantities satisfy a linear Smarr relation
    \begin{align}
    	M=2TS+\Psi_{H}Q_{m}+2K_{\alpha}\alpha-2PV\ ,
    \end{align}
     where $M$ is again the Komar mass. We will now rewrite the first law and Smarr relation solely in terms of the physically relevant parameters i.e. the horizon area $A$, the minimal length $l$ and the cosmological constant $\Lambda$, instead of the parameters $(Q_{m},\alpha,P)$ that appear in \eqref{eq:1st law-gen-hayward}. We have that
    \begin{align}
    	Q_{m}=\frac{Al^{1/3}}{8\pi}\left(\frac{12\pi+8\pi A P}{3A-12\pi l^2-8\pi Al^2 P}\right)\ ,
    \end{align} 
    and 
    \begin{align}
    	\alpha=2l^2\ ,
    \end{align}
    with their differentials being given by
    \begin{align}
    	dQ_{m}=\frac{\partial Q_{m}}{\partial A}dA+\frac{\partial Q_{m}}{\partial l}dl+\frac{\partial Q_{m}}{\partial P}dP\ ,
    \end{align}
    \begin{align}
    	d\alpha=\frac{\partial \alpha}{\partial A}dA+\frac{\partial \alpha}{\partial l}dl+\frac{\partial \alpha}{\partial P}d P\ .
    \end{align}
   This gives the following form of modified first law
    \begin{align}\label{tildefirstlaw}
    	dM=\tilde{T}dS+\tilde{\Phi}dl+\tilde{V}dP\ ,
    \end{align}
    where an effective temperature $\tilde{T}$ given by
    \begin{align}
    	\tilde{T}=T+4\Psi_{H}\frac{\partial Q_{m}}{\partial A},
    \end{align}
   naturally emerges, along with an effective potential $\tilde{\Phi}$ associated with the minimal length
    \begin{align}
    	\tilde{\Phi}=\Psi_{H}\frac{\partial Q_{m}}{\partial l}+K_{\alpha}\frac{\partial \alpha}{\partial l}\ ,
    \end{align}
    and an effective thermodynamic volume $\tilde{V}$
    \begin{align}
    	\tilde{V}=V+\Psi_{H}\frac{\partial Q_{m}}{\partial P}\ ,
    \end{align}
    where $V=4\pi r^{3}_{h}/3$. These effective quantities also define a proper Smarr relation 
    \begin{align}
    	M=2\tilde{T}S+\tilde{\Phi}l-2\tilde{V}P\ ,
    \end{align} 
    with the appropriate scaling invariance \cite{kastor2009} under
    \begin{align}
    	M\rightarrow cM,\quad l\rightarrow cl,\quad r_h\rightarrow cr_h,\quad P\rightarrow c^{-2}P,\nonumber\\ 
       	S\rightarrow cS^2,\quad \tilde{T}\rightarrow c^{-1}\tilde{T},\quad \tilde{\Phi}\rightarrow \tilde{\Phi},\quad \tilde{V}\rightarrow c^{3}\tilde{V}\ .
    \end{align}
    Linearity of the Smarr relation is intimately tied to the symmetry inheritance properties of the gauge fields present in the theory. It is often assumed that if the underlying metric $g$ possesses a symmetry $\mathcal{L}_\xi g_{ab}=0$ generated by a Killing field $\xi$, the electromagnetic field will inherit the same symmetry, i.e. $\mathcal{L}_\xi F_{ab}=0$. In $d=4$ Einstein-Maxwell theory there are known counterexamples to this assumption, as seen in certain classes of Bianchi type-V metrics \cite{ftaclas1978} and pp-wave metrics \cite{wainwright1976}, a phenomenon which extends to other symmetries such as the conformal group in multifluid spacetimes \cite{coley1990}. The symmetry inheritance property of the field is important for establishing constancy of the electromagnetic potential on the Killing horizon, and is also required for establishing linearity of the Smarr relation, as was done in \cite{gulin2018}. In our case the gauge field does {\it not} inherit the symmetry of the spacetime, with $\mathcal{L}_{\xi_{\theta}}g=0$ but $\mathcal{L}_{\xi_{\theta}}A^{\mu}\neq0$, yet the Smarr relation remains linear. This suggests the condition presented in \cite{gulin2018} may be sufficient but not necessary for linearity of the Smarr relation.
\\

    The explicit form of the quantities appearing in \eqref{tildefirstlaw} are given below. The effective temperature is
    \begin{align}
    	\tilde{T}=\frac{3r_h\left(3r^2_h-3r^4_h\Lambda-l^2(-3+r^2_h\Lambda)^2\right)}{4\pi\left(3r^2_h+l^2(-3+r^2_h\Lambda)\right)^2}\ ,\label{eq:eff-temp}
    \end{align}
    which is evidently different from the temperature \eqref{eq:temp-surf.grav} defined through the surface gravity, though they are simply related through
    \begin{align}
    	\tilde{T}=\frac{T}{\left(1-\frac{l^2}{r^2_h}\left(1-\frac{\Lambda}{3}r^2_h\right)\right)^2}\ .
    \end{align}
    The difference between the two arises from how one treats the contribution to the total energy variation of the first law arising from the matter, which may or may not be thermalized with respect to the Killing time. When matter is present, the entropy variation conjugate to the surface gravity $\kappa$ does not necessarily refer only to the geometric entropy of the black hole, and requiring it to do so is accompanied by a modification of the conjugate temperature. The root cause arises from an ambiguity in the splitting of the Lagrangian into a geometric and matter contribution, which is manifest in the covariant formalism \cite{iyer1997}. A similar modification can occur, for example, when scalar fields are present which do not minimally couple to gravity, as recently observed in $d=4$ Gauss-Bonnet gravity \cite{LHK:23}.  When the minimal length vanishes however, the effective temperature and the surface gravity temperature coincide, as expected since in this limit the matter contribution to the total Lagrangian vanishes. The two remaining potentials are given by
    \begin{align}
    	\tilde{\Phi}=\frac{lr^3_h(-3+\Lambda r^2_h)^2}{\left(3r^2_h+l^2(-3+\Lambda r^2_h)\right)^2}\ ,\label{eq:phi-tilde}
    \end{align}
    \begin{align}
    	\tilde{V}=\frac{12r^7_h}{\left(3r^2_h+l^2(-3+\Lambda r^2_h)\right)^2}\ .\label{eq:V-tilde}
    \end{align}
    Note that the same version of the first law \eqref{tildefirstlaw} is obtained if one uses the fact that $Q_m=Q_m(m,l)$ and $\alpha=\alpha(l)$ through \eqref{eq:Qm} and \eqref{eq:alpha-q}, giving
    \begin{align}
    dM&=TdS+VdP+\Psi_{H}dQ_{m}+K_{\alpha}d\alpha\nonumber\\
    &=TdS+VdP+\left(\Psi_{H}\frac{\partial Q_m}{\partial m}\right) dm\\
    &\qquad\qquad\qquad\qquad+\left(\frac{\partial Q_m}{\partial l} +K_{\alpha}\frac{\partial \alpha}{\partial l}\right)dl\ .\nonumber
    \end{align}
    Now observing that for the Hayward metric, $m$ is in fact equal to the ADM/Komar mass $M$ appearing on the left-hand side of \eqref{eq:1st law-gen-hayward}, one has that the variation of the mass satisfies
    \begin{align}
    \begin{aligned}
    dM=&\left(\frac{T}{1-\Psi_{H}\frac{\partial Q_m}{\partial m}}\right)dS+\left(\dfrac{V}{1-\Psi_{H}\frac{\partial Q_m}{\partial m}}\right)dP\\
    &+\left(\dfrac{\frac{\partial Q_m}{\partial l} +K_{\alpha}\frac{\partial \alpha}{\partial l}}{1-\Psi_{H}\frac{\partial Q_m}{\partial m}}\right)dl.
    \end{aligned}
    \end{align}
    where the quantities in brackets will be the same as the effective quantities appearing in \eqref{tildefirstlaw}.
    \subsection{Euclidean path integral methods}\label{euclidean}
     We will now elaborate on the relationship between the thermodynamic variables obtained from the Hamiltonian version of the first law above and another commonly employed method for understanding black hole thermodynamic---the Euclidean path integral. Developed by Gibbons and Hawking \cite{gibbons1977} and extended by York \cite{york1986,braden1990}, this method has a basis in the fundamental relationship between the partition function $\mathcal{Z}$ of general quantum systems and the Euclidean path integral. The partition function for a continuous quantum system defined by canonical variables $\{q_i\}$ with Hamiltonian $H$ at finite temperature $T=\beta^{-1}$ is 
    \be\label{euclidean1}
    \mathcal{Z}=\int dq_i \bra{q_i}e^{-\beta H}\ket{q_i}\ ,\nonumber
    \ee
    from which thermodynamic quantities for the statistical ensemble can be readily determined as
    \be\label{thermo1}
    F=-T \ln \mathcal{Z}\ ,\quad E=-\dfrac{\partial \ln\mathcal{Z}}{\partial \beta}\ ,\quad S=-\beta\dfrac{\partial \ln\mathcal{Z}}{\partial \beta}+\ln\mathcal{Z}\ .
    \ee	
    In the context of gravity, $\mathcal{Z}$ is formally given by an intractable path integral over an ill-defined measure $\mathcal{D}[g]$ which includes contributions from both matter and gravitational degrees of freedom. The partition function can however be computed in a semiclassical saddle-point approximation where
    \be
    \mathcal{Z}=\int_{g(0)}^{g(\tau)}\!\! \mathcal{D}[g]\ e^{-I_E[g]}\approx\sum_{g_{cl}} e^{-I_E\left[g_{c l}\right]}\ ,\nonumber
    \ee
    where $I_E[g]$ is the Euclidean action of the metric $g$, and $I_E\left[g_{c l}\right]$ is the saddle-point contribution from the Euclidean metrics $g_{cl}$ which solve the classical equations of motion and obey the prescribed boundary conditions\footnote{As discussed in \cite{Banihashemi2022}, there are subtleties involved in this approximation which should be treated carefully.}. The periodicity in $\tau$ implements the trace from \eqref{euclidean1} and encodes the Kubo-Martin-Schwinger (KMS) condition for finite-temperature fields \cite{kubo1957,martin1959}. We work in an on-shell spherical reduction, where the periodicity is naturally fixed to remove a would-be conical singularity at the origin of the Euclidean section (the black hole event horizon):
    \be
    \beta^{-1}=\dfrac{\kappa}{2\pi}.
    \ee	
    This method of computing thermodynamic quantities is particularly advantageous for black holes in de Sitter space, because one can fix boundary-value data (in our case the temperature) on a surface at some finite radius $r_c$ between the black hole and cosmological horizons, and compute $\mathcal{Z}$ by performing the direct (now finite) integration. This physically corresponds to placing the black hole in an isothermal cavity, where the temperature at the cavity is fixed to be 
    \be
    \beta_c^{-1}=\dfrac{\kappa}{2\pi \sqrt{f(r_c)}}\ ,
    \ee
    which is just the locally observed KMS temperature at $r_c$. This method has been applied to a wide variety of black hole spacetimes where a thermodynamic landscape comparably rich to that of AdS has been revealed \cite{carlip2003,simovic2019,haroon2020,simovic2021}. In the case of the Hayward black hole model considered in this work we require the total reduced Euclidean action $I_{r}$ for the Einstein-Hilbert-NED theory, given by
    \begin{align}
    	I_{r}=I_{EH}+I_{GHY}+I_{M}+I_{EMB}-I_{0},
    \end{align}
    where $I_{EH}$ is the Einstein-Hilbert action, $I_{GHY}$ the Gibbons-Hawking-York boundary term, $I_{M}$ is the Euclidean action for any matter fields present, $I_{EMB}$ is an electromagnetic boundary term required to fix the charge, and $I_{0}$ is a subtraction term which serves to regularize the infinite volume integral for a spacetime without boundary. Explicit calculation of each term can be found in the Appendix \ref{sec:app:eucl-action}. The individual terms are defined as
    \begin{align}
    	I_{EH}=-\frac{1}{16\pi}\int_{\mathcal{M}} d^4x\sqrt{g}\left(R-2\Lambda\right)\ ,
    \end{align} 	
    \begin{align}
    	I_{GHY}=\frac{1}{8\pi}\int_{\partial \mathcal{M}}d^3x \sqrt{k}K\ ,\label{eq:Ighy}
    \end{align}
    where $K$ is the trace of the extrinsic curvature of the spherical boundary $\partial \mathcal{M}$  at $r_c$ and $k$ is the determinant of the boundary metric,
    \begin{align}
    	I_{M}=\frac{1}{16\pi} \int_{\mathcal{M}} d^4x \sqrt{g}\mathcal{L}(\mathcal{F}) ,\label{eq:Imatter}
    \end{align}
    is the matter action,
    \be
    I_{EMB}=-\frac{1}{16\pi}\int_{\partial \mathcal{M}}d^3x \sqrt{k}\ \left(\frac{\partial \mathcal{L}}{\partial \mathcal{F}}\right)F^{\mu\nu}n_{\nu}A_{\mu}\ ,
    \ee
    is an electromagnetic boundary term which fixes the charge in the canonical ensemble, and the subtraction term $I_{0}$ is simply the total reduced action evaluated for the empty spacetime (Minkowski, AdS, or dS space, depending on which asymptotics are assumed). In asymptotically flat or AdS space this term removes the divergent part of the volume integral. In the case where the integration is cut at a finite boundary no such divergence appears, and instead $I_0$ simply normalizes the action such that $I_r=0$ for the empty spacetime. After combining the terms and performing the required integrations, we find that total reduced action is given by  
    \begin{align}
    	I_{r}=&-\pi r^2_h+\frac{\beta r_c}{3} \left(3-r^2_c\Lambda -\sqrt{3-\Lambda r^2_c}\,\mathcal{X}\right)\ ,\label{eq:red-action}
    \end{align}
    where
    \be
    \mathcal{X}\equiv \sqrt{\frac{(3-\Lambda r^2_c)\mathcal{Y}-9\,r^2_c\, r^2_h\big(r_c-r_h-\tfrac{\Lambda}{3}(r^3_c +r^3_h)\big)}{\mathcal{Y}-3\,r^3_c\, r^2_h}},
    \ee
    and 
    \be
    \mathcal{Y} \equiv l^2(r^3_c-r^3_h)(r^2_h\Lambda-3)\ .
    \ee
    The $\beta$-independent term ensures that the correct entropy is obtained from Eq.\eqref{thermo1} which reduces to, 
    \begin{align}
    	S=\beta \frac{\partial I_{r}}{\partial \beta}-I_{r}=\pi r^2_h.
    \end{align}
    The equilibrium temperature can be obtained by extremizing the action with respect to $r_{h}$ and solving for $\beta$, 
    \begin{align}
    	\frac{\partial I_{r}}{\partial r_{h}}=0\ ,
    \end{align} 
    which gives a temperature $\mathcal{T}=\beta^{-1}$ of
    \begin{align}
    	\mathcal{T}=\frac{3r^6_c\sqrt{3-r^2_c\Lambda}\left(3r^3_h(1-r^2_h\Lambda)-l^2r_h(r^2_h\Lambda-3)^2\right)}{4\pi \big(3\,r^3_c\,r^2_h+\mathcal{Y})\big)^2\mathcal{X}}\ .\label{eq:temp-eucl}
    \end{align}
    If there is no cosmological horizon then one can take the limit $r_c\rightarrow \infty$ and the above relation for  $\mathcal{T}$ reduces to the effective temperature $\tilde{T}$ defined in Eq. \eqref{eq:eff-temp}. The mean thermal energy is 
    \be
    E=\dfrac{\partial I_r}{\partial\beta}=\frac{r_c}{3} \left(3-r^2_c\Lambda -\sqrt{3-\Lambda r^2_c}\,\mathcal{X}\right)\ ,\label{eq:energy-eucl}
    \ee
    and the potentials conjugate to $l$ and $\Lambda$ can be obtained by the variations
    \begin{align}
    	\varphi=\frac{1}{\beta}\frac{\partial I_{r}}{\partial l}\ ,\quad \mathcal{V}=-\frac{8\pi}{\beta}\frac{\partial I_{r}}{\partial \Lambda}\ ,
    \end{align}
    which give 
    \begin{align}
    	\varphi=\frac{l r^3_cr^3_h(r^3_c-r^3_h)\sqrt{3-r^2_c\Lambda}(-3+r^2_h\Lambda)^2}{ \left(3r^3_cr^2_h+l^2(r^3_c-r^3_h)(-3+r^2_h\Lambda)\right)^2\mathcal{X}},\label{eq:phi-eucl}
    \end{align}
    and
    \begin{align}
    	\mathcal{V}&=\frac{8\pi r_c^3}{3} \left(1+\frac{\sqrt{3-\Lambda  r_c^2}\ \mathcal{X}'}{r_c^2}-\frac{\mathcal{X}}{2 \sqrt{3-\Lambda  r_c^2}}\right)\ ,\label{eq:V-eucl}
    \end{align}
    where a prime denotes a derivative with respect to $\Lambda$. For asymptotically flat ($\Lambda=0$) or anti-de Sitter $(\Lambda<0$) spacetimes, one can take the limit $r_{c}\rightarrow \infty$ and find that $\varphi$ and $\mathcal{V}$ coincide with $\tilde{\Phi}$ and $\tilde{V}$, which are respectively given by Eqs. \eqref{eq:phi-tilde} and \eqref{eq:V-tilde}, demonstrating that the Euclidean formulation is consistent with the Hamiltonian one used in the AdS case, in the regime where the two methods are comparable. The Euclidean formulation is however more broadly applicable since it can be implemented for any standard choice of asymptotics, while the Hamiltonian formulation requires a Killing vector field which is timelike in the asymptotic region to associate the conserved energy with.
    \\
    
    In the presence of a cavity, and additional work term $\lambda dA_c$ enters into the first law relating the cavity tension $\lambda$ and variations of its area $A_c$ \cite{braden1990}. The tension $\lambda$ is determined through
    \begin{align}
    	\lambda=\frac{1}{\beta}\frac{\partial  I_{r}}{\partial A_c}\ ,
    \end{align}
    and is given in Appendix \ref{sec:app:lambda}. With the quantities above, the following form of the Smarr relation is satisfied
    \begin{align}
    	E=2\mathcal{T}S+\varphi l+2\lambda A_c-2P\mathcal{V}\ .
    \end{align}
    Note that the thermal energy $E$ and the mass $M$ are not equal except in the limit when $r_c\rightarrow \infty$.  We now argue that the correct choice of thermodynamic variables entering the first law for the Hayward solution are the ones given above, based on their agreement with the quantities computed from the Hamiltonian construction (when appropriate limits can be taken to sensibly relate the two). As there are a number of different frameworks available for computing thermodynamic quantities associated with black hole spacetimes, a natural question is whether one should expect the Euclidean path integral method of Gibbons and Hawking to agree with a Hamiltonian formulation in the first place. This has been discussed previously by Iyer and Wald \cite{iyer1995a}, who showed that the covariant phase space construction of the first law (under which the Hamiltonian form originally given by \cite{BCH:73} is subsumed) is equivalent to a Euclidean path integral formulation based on the semiclassical approximation to the partition function, provided that the Lagrangian is at most linear in the curvature.
    
    \section{Hayward phase structure} \label{sec:phase-structure}
    \subsection{Anti-de Sitter $\Lambda<0$}\label{ads}
    Asymptotically anti-de Sitter black holes readily admit a thermodynamic interpretation owing to the confining effective potential of anti-de Sitter space. Free particles (both massive and massless) follow periodic closed orbits in an AdS background, so Hawking radiation from a black hole is naturally confined provided reflecting boundary conditions are imposed at the timelike boundary. Therefore, subtleties in determining when/if thermodynamic equilibrium is achieved by a large black hole are avoided since one can simply assume that sufficient time has passed for the outgoing radiation to equilibrate with the black hole (provided it is sufficiently large compared to the AdS length scale). It has long been known that the unique boundary conditions of anti-de Sitter space allow for a gauge-gravity duality relating quantum gravity theory in the bulk of AdS to a conformal field theory on the boundary of AdS. This is the basis of the anti-de Sitter/conformal field theory correspondence, which has proven to be both extremely useful in practical computations and of great theoretical importance. The prototypical example is the exact correspondence between type IIB string theory in the bulk and $\mathcal{N}=4$ super-Yang Mills theory on the boundary \cite{maldacena1999}. The study of bulk gravitational physics in asymptotically AdS spacetimes is therefore of direct relevance for the study of strongly coupled gauge theories \cite{son2002}.
    \\
    
    It is in this asymptotically anti-de Sitter context that the well-known Hawking-Page \cite{HP:83} transition occurs: at low temperature a would-be small black hole will simply evaporate, and the AdS space is eventually filled with free radiation. A sufficiently large black hole however will persist long enough to come into equilibrium with its own Hawking radiation, which returns from the boundary in finite time. There is a critical temperature $T_c$ separating these two regimes at which the Hawking-Page transition occurs. Note that unlike asymptotically flat black holes, the temperature of asymptotically AdS black holes does not monotonically decrease with the size of the black hole. This Hawking-Page transition corresponds to a deconfinement transition in the boundary CFT, and has long been studied to understand nonperturbative features of strongly coupled CFT systems and the black hole information problem \cite{witten1998,harlow2016}.
    \\
    
\begin{figure}[h]
    	\centering
    	\includegraphics[width=0.485\textwidth]{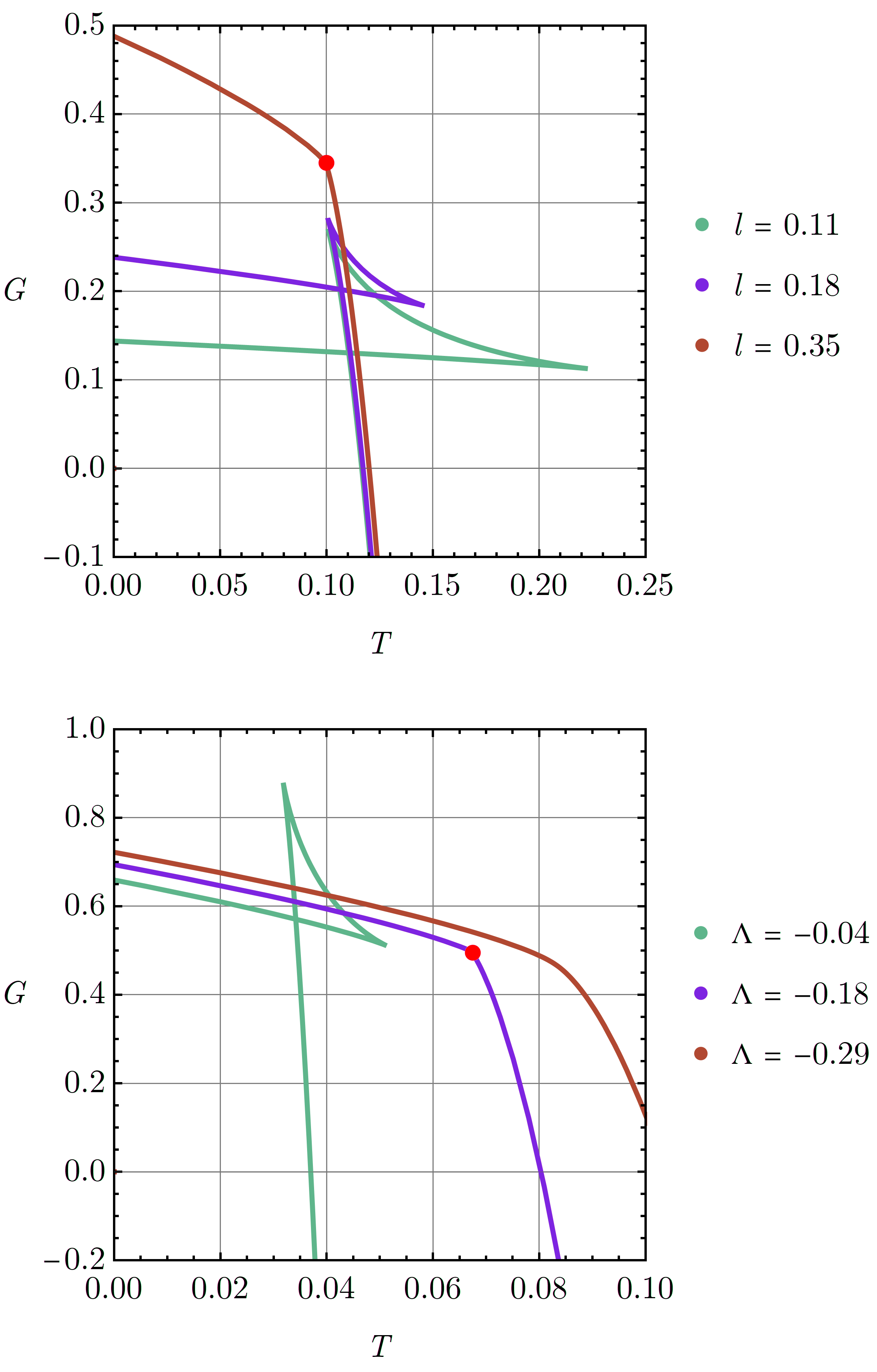}
    	\caption{\small{Gibbs free energy $G$ as a function of temperature $T$ for the Hayward-AdS black hole. The horizon size $r_h$ increases in the direction of the arrows. {\bf Top:} Fixed $\Lambda=-0.04$ for various $l$. The onset of a first-order phase transition from a small to large black hole is marked by a red dot at the critical temperature $T_c\sim 0.1$. {\bf Bottom:} Fixed $l=0.5$ for various $\Lambda$. The critical temperature occurs at $T_c\sim 0.68$.}}
    	\label{GT1}
    \end{figure}

    We would like to examine whether this transition persists when the central singularity is absent in the classical metric, and whether new types of transitions appear. The phase structure of the Hayward-AdS black hole can be determined through the Gibbs free energy $G=M-\tilde{T}S$, which is globally minimized by the equilibrium state of the system. Using \eqref{eq:mass} and \eqref{eq:eff-temp}, the free energy is
    \be
    G=\frac{l^2 r_h^3 \left(\Lambda r_h^2-3\right)^2+3 r_h^5 \left(\Lambda r_h^2+3\right)}{4 \left(l^2 \left(\Lambda r_h^2-3\right)+3 r_h^2\right)^2}\ ,
    \ee
    which can be plotted parametrically as a function of the equilibrium temperature $\tilde{T}$ using $r_h$ as a parameter. The result is shown in Fig.~\ref{GT1}.\\
    
    Two characteristically distinct behaviors are revealed in Fig.~\ref{GT1} depending on the value of the cosmological constant $\Lambda$ and the regularization length scale $l$. For fixed $\Lambda=-0.04$, there is a critical length scale $l_c\approx 0.35$ below which a first-order phase transition from a small to large black hole occurs. Likewise when $l$ is fixed, there is a critical $\Lambda_c$ below which a similar transition is present. Above both critical values there is no phase transition, while at the critical point the transition becomes second-order (the precise location of the critical points for variable $l$ and $\Lambda$ is determined below). This is reminiscent of the small-large transition observed previously in Reissner-Nordstrom-AdS black holes \cite{carlip2003}. Although the small black hole phase has greater free energy than the radiation phase at $G=0$, the would-be Hawking-Page transition does not occur since the system is in a fixed-charge (canonical) ensemble. Unlike the Reissner-Nordstrom case however, the presence of charge here is tied directly to the existence of a minimal length scale. The radiation phase is inaccessible as long as $Q_m>0$ which through \eqref{eq:alpha-q} corresponds to $l>0$. Therefore, the presence of a quantum gravity regulator of the Schwarzschild singularity appears to prevent the Hawking-Page transition from occurring. 
    \\
    
    This has important implications for AdS-CFT, because it is usually assumed that the bulk gravity theory is some suitable low-energy limit of a full quantum gravity theory. The standard Hawking-Page transition is between a classical Schwarzschild black hole and the empty AdS geometry, corresponding to a deconfinement transition in the dual CFT. What we demonstrate is that a first-order approximation to a regularized Schwarzschild geometry, where there is no backreaction and the action does not yet include higher curvature terms in an effective field theory description of the full gravitational sector, eliminates the Hawking-Page transition altogether, at least when the regulator can be understood through the effective action of a $U(1)$ gauge field. At the same time it is known that the Hawking-Page transition persists when higher curvature corrections like $R^2-4R_{ab}R^{ab}+R_{abcd}R^{abcd}$ are explicitly included in the bulk gravitational action, though such corrections do not regularize the Schwarzschild singularity. Since the Hayward metric is a rather generic approximation to a smoothed geometry which can in principle arise from any quantum theory of gravity, we propose that quantum gravity corrections to Einstein-Hilbert gravity which are in some sense ``strong enough" to regularize the central singularity may generically prevent the Hawking-Page transition from occurring, and likewise will have important implications for the critical behavior of the dual CFT theory.
    \\
    
\begin{figure}[h]
    	\centering
    	\includegraphics[width=0.48\textwidth]{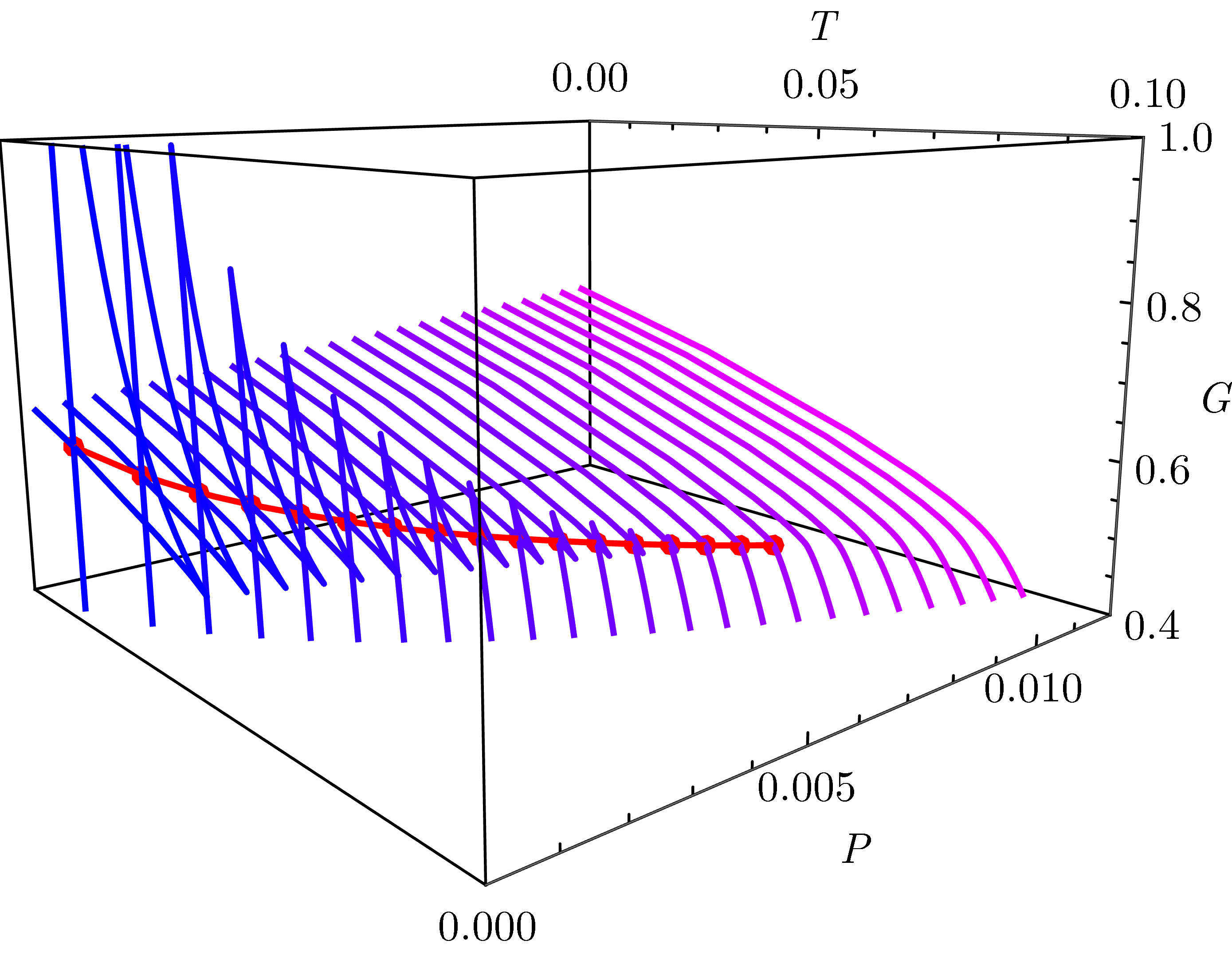}
    	\caption{\small{Gibbs free energy $G$ as a function of temperature $T$ and pressure $P$ for the Hayward-AdS black hole, demonstrating the formation of a swallowtail below the critical point. For $l=0.5$ this occurs as $\{P_c,T_c\}=\{0.0083,0.072\}$.}}
    	\label{tail}
    \end{figure}

\begin{figure}[h]
    	\centering
    	\includegraphics[width=0.38\textwidth]{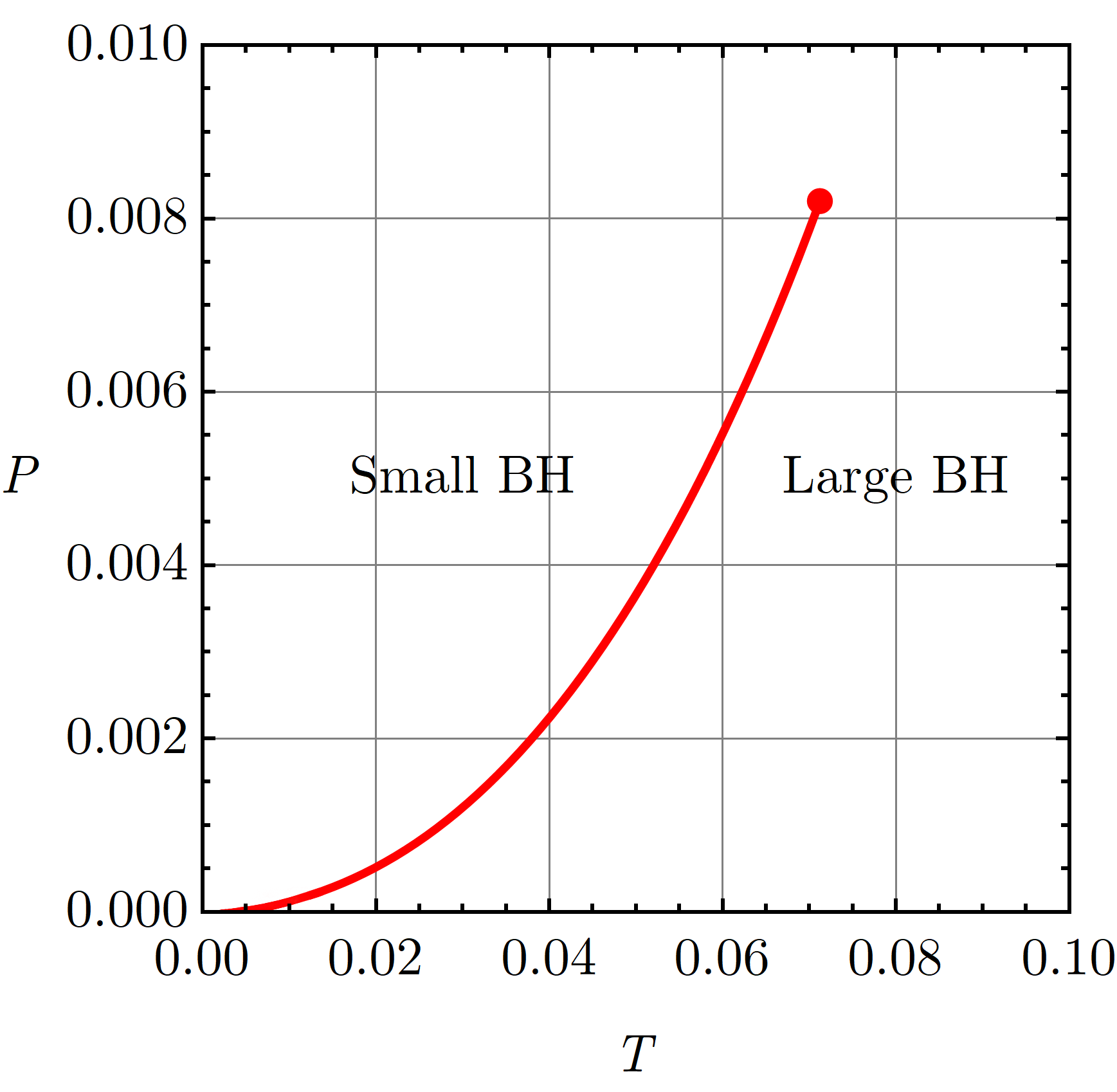}\hspace{10pt}
    	\caption{\small{Coexistence line for the Hayward-AdS black hole for $l=0.5$. A series of first-order phase transitions occur along the red line, terminating at a critical point $\{P_c,T_c\}= \{0.0083,0.072\}$ where the transition becomes second-order.}}
    	\label{coex1}
    \end{figure}

\begin{figure}[h]
    	\centering
    	\includegraphics[width=0.49\textwidth]{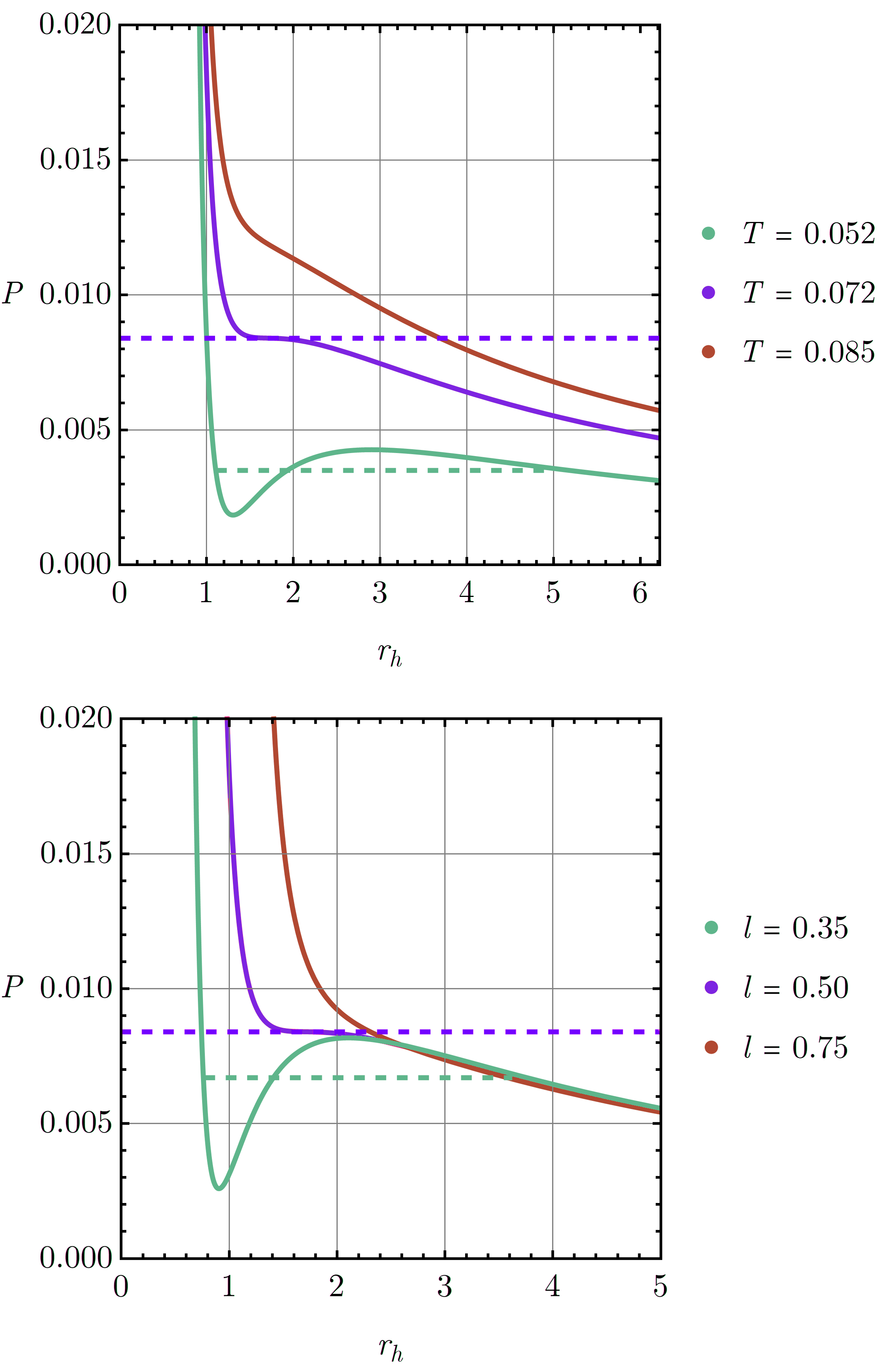}
    	\caption{\small{Isotherms of the equation of state $P(r_h,T)$ for the Hayward-AdS black hole. {\bf Top:} Fixed $l=0.5$ for various $T$. The onset of the Van der Waals transition is marked by an inflection point at the critical temperature $T_c\sim 0.072$. {\bf Bottom:} Fixed $T=0.072$ with varying $l$. The critical pressure is marked by the red dashed line, while the coexistence line is marked by the blue dashed line.}}
    	\label{PV1}
    \end{figure}

    In the extended phase space, the Hayward-AdS black hole exhibits critical behavior analogous to the mean-field theory critical behavior of ordinary fluid systems (see \cite{KMT:17} for a review). In Fig.~\ref{tail} we demonstrate the standard ``swallowtail" behavior that generically occurs for asymptotically AdS black holes in the extended phase space, with the corresponding coexistence line shown in Fig.~\ref{coex1}. Above the critical pressure (critical $\Lambda$) is a ``superfluid" phase where the system smoothly transitions from a small to a large black hole as the temperature is increased. This transition is exactly analogous to the liquid-gas transition occurring in a traditional van der Waals fluid system. The transition can be examined further by constructing the equation of state $P(V,T)$ for the system. With $\Lambda=-8\pi P$ one obtains from \eqref{eq:eff-temp} that
    \be
    P=\frac{3 \left(3r_h^5-2 l^2 r_h^3 (3-4 \pi  r_h T)-8 \pi  l^4 r_h^2 T-\sqrt{r_h^7 X}\right)}{16 \pi  l^2 r_h^4 \left(4 \pi  l^2 T+3 r_h\right)}\ ,\label{eq:pressure}
    \ee
    where $X\equiv 9r_h^3-24 l^2 r_h-32 \pi  l^4 T$. Here and in what follows, we omit the tilde on thermodynamic variables with the understanding that they refer to the quantities in \eqref{tildefirstlaw}. The thermodynamic volume is a 7th degree polynomial in $r_h$,
    \be
    V=\frac{12 \pi r_h^7}{ \left(3 r_h^2+l^2 \left(\Lambda r_h^2-3\right)\right)^2}\ ,
    \ee
    preventing one from writing the equation of state in analytic form as $P(V,T)$. However since the volume $V(r_h,l)$ is a monotonic function of the horizon size $r_h$ for fixed length scale $l$, one can freely use $r_h$ as a parameter. In Fig.~\ref{PV1} the equation of state is given as a function of horizon size $r_h$ (it is is qualitatively identical when given parametrically as a function of $V$). The red dashed line marks the onset of the small-large transition at the critical point, which simultaneously satisfies
    \be\label{crit1}
    \dfrac{\partial P}{\partial V}=0\ ,\qquad \dfrac{\partial^2 P}{\partial V^2}=0\ .
    \ee

    Below the critical temperature/length scale there is an unphysical region where $\partial P/\partial V>0$, and two more unstable regions. One may also observe that if the temperature $T$ or critical length scale $l$ are small enough, the system appears to enter a region where $P<0$ corresponding to a transition to asymptotically de Sitter space. Both pathologies are avoided because the system instead evolves along the dashed line of Fig.~\ref{PV1}, which is determined by the condition that the small and large black hole phases have the same free energy. This coexistence line always lies above $P=0$. Fig.~\ref{GT1} indicates a minimal horizon size at $T=0$, which corresponds to the extremal limit and implies the constraint that 
    \begin{align}\label{constraint1}
    r_h>\sqrt{\frac{3+6l^2\Lambda-3\sqrt{1-8l^2\Lambda}}{6\Lambda+2l^2\Lambda^2}}\ .
    \end{align}
    On the other hand, at fixed $T>0$ the equation of state instead implies a lower bound of 
    \be\label{constraint2}
    r_h>\frac{ \sqrt[3]{16} \left[\left(l^3\sqrt{ 9 \pi ^2 l^2 T^2-2}+3 \pi  l^4 T\right)^{2/3}+\sqrt[3]{2} l^2\right]}{3 \sqrt[3]{l^3\sqrt{ 9 \pi ^2 l^2 T^2-2}+3 \pi  l^4 T}}\ ,
    \ee
    below which the pressure becomes imaginary. Surprisingly, the critical condition \eqref{crit1} can be solved analytically to find
    \begin{align}
    P_c&= \dfrac{5\sqrt{10}-13}{432\pi l^2}\ ,\quad V_c= \frac{8\pi \sqrt{\frac{2}{5}} \left(5 \sqrt{10}+13\right)^{7/2}   l^3}{375 \left(\sqrt{10}+2\right)^2}\ ,\ \nonumber\\  &\qquad\qquad T_c=\frac{\sqrt{13 \sqrt{\frac{5}{2}}-\frac{31}{2}}}{20 \pi  l}\ .\label{crit}
    \end{align}
    In terms of the reduced volume $v=2 r_h l_p^2$ the `universal' ratio $P_c v_c/T_c$ is
    \be
    \dfrac{P_c v_c}{T_c}\approx0.393\neq 3/8\ ,\label{eq:crit-ratio}
    \ee
    which represents a rare counterexample of the usual $3/8$ result for a four-dimensional black hole spacetime, signaling a potential departure from mean field theory critical behavior. This same ratio was computed in \cite{fan2017} using $\alpha$ and $Q_m$ as the fundamental variable parameters of the theory. However we believe their analysis contained some inaccuracies which inevitably lead to different critical exponents, as detailed below. 
    \\
    
    In previous work \cite{KMT:17}, the reduced volume $v$ was identified with the horizon radius by comparing the linear$-T$ coefficient of the black hole equation of state to that of the van der Waals fluid. One can proceed the same way here by expanding Eq. \eqref{eq:pressure} in powers of $T$ to obtain
    \begin{align}
    	P=P_{0}+\frac{T}{\tilde{v}}+\mathcal{O}(T^2),
    \end{align}    
    where 
    \begin{align}
    	P_{0}=\frac{3r^2_h-6l^2-r_h\sqrt{9r^2_h-24l^2}}{16\pi l^2r^2_h},
    \end{align}
    and the new reduced volume $\tilde{v}$ is identified to be
    \begin{align}
    	\tilde{v}=\frac{12 r^2_h(3r^2_h-8l^2)}{(3r^2_h-4l^2)\sqrt{9r^2_h-24l^2}-24l^2r_h+9r^3_h}\ .\label{eq:eff-red-volume}
    \end{align}
    In the limit $l\rightarrow 0$ the reduced volume $\tilde{v}$ reduces to $\tilde{v}=2r_{h}+\mathcal{O}(l^2)$. Evaluating the critical ratio $P_c \tilde{v}_c/T_c$ in terms of this new reduced volume, we find that
    \begin{align}
    	\frac{P_{c}\tilde{v}_{c}}{T_{c}}\approx 0.391 \neq \frac{3}{8},
    \end{align}
    which is a small deviation from the value \eqref{eq:crit-ratio} obtained using the conventional reduced volume $v=2r_{h}$. In both cases the ratio differs from $3/8$.
    \\
    
\begin{figure}[h]
    	\centering
    	\includegraphics[width=0.48\textwidth]{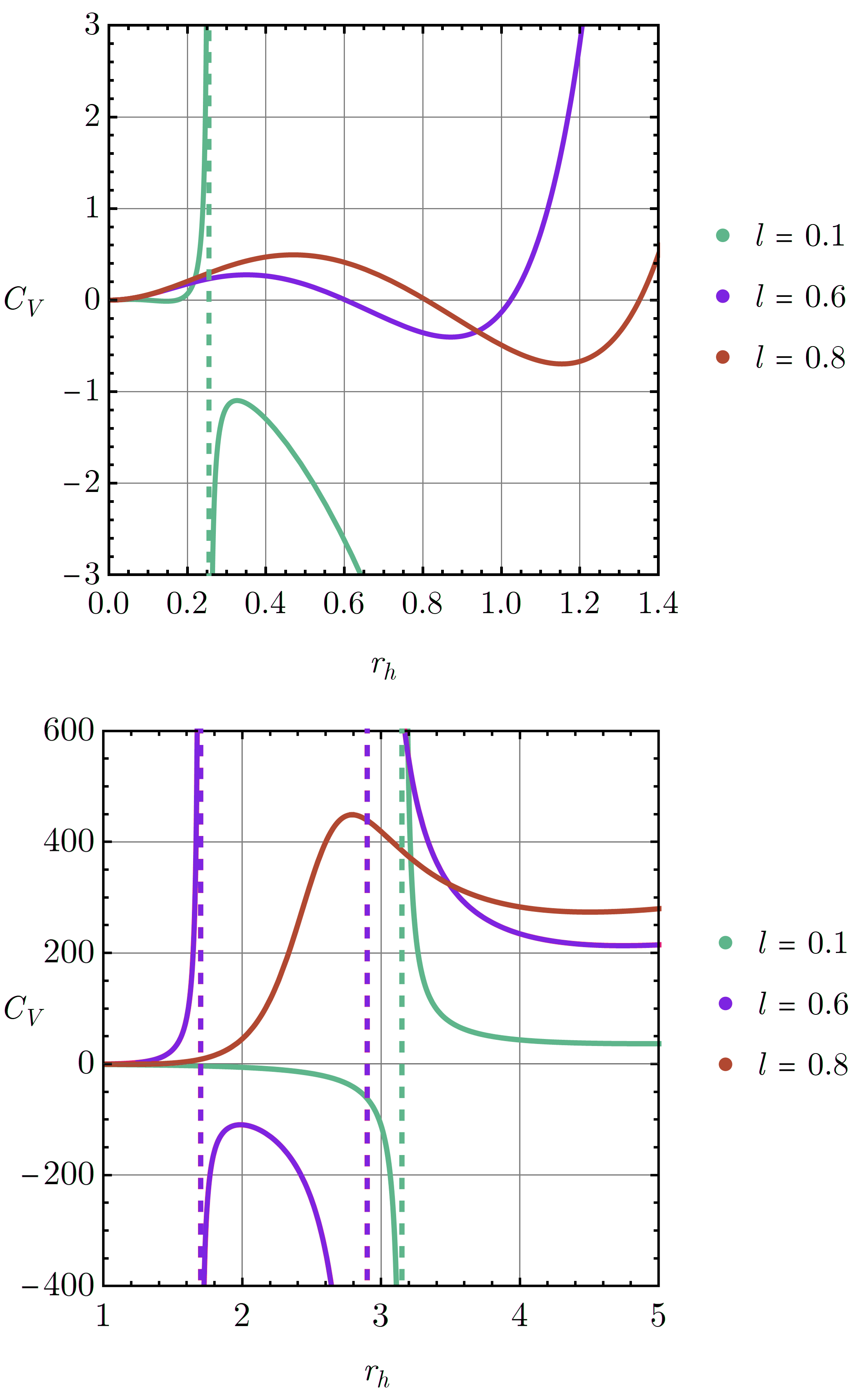}
    	\caption{\small{Heat capacity $C_V$ as a function of horizon size $r_h$ for the Hayward-AdS black hole, for $\Lambda=-0.1$ and various $l$. Dashed lines correspond to a discontinuity in $C_V$. {\bf Top:} Small-$r_h$ behavior. {\bf Bottom:} Large-$r_h$ behavior.}}
    	\label{cv1}
    \end{figure}

     Thermodynamic stability can further be assessed using the heat capacity, which is readily determined to be
    \begin{align}\label{cv}
    &C_V=T\left(\dfrac{\partial S}{\partial T}\right)\bigg|_V\\
    &\!\!\!=\frac{2 \pi  r_h^2 \left(l^2\! \left(\Lambda r_h^2\!-\!3\right)\!+\!3 r_h^2\right) \left(l^2\! \left(\Lambda r_h^2\!-\!3\right)^2\!+3 \Lambda r_h^4\!-\!3 r_h^2\right)}{l^4\! \left(\Lambda r_h^2-3\right)^3\!+6 l^2 r_h^2 \left(\Lambda r_h^2 \left(\Lambda r_h^2\!-\!4\right)\!-\!9\right)\!+\!9 \left(\Lambda r_h^6+r_h^4\right)},\nonumber
    \end{align}
    and is plotted in Fig.~\ref{cv1} for varying $l$. Positivity of the heat capacity indicates thermodynamic stability, while a divergence in the heat capacity generically indicates a phase transition. A sign change in $C_V$ can occur either through a discontinuity or zero-crossing, with the divergences indicated by a vertical dashed line in the figure. Note that some of the unstable regions are excluded from the parameter space since they do not satisfy the constraint \eqref{constraint2}. The heat capacity diverges when
    \begin{align}
    &r_h^6 \left(l^4 \Lambda^3+6 l^2 \Lambda^2+9 \Lambda\right)+r_h^4\left(-9 l^4 \Lambda^2-24 l^2 \Lambda+9\right)\nonumber\\
    &+r_h^2 \left(27 l^4 \Lambda-54 l^2\right)-27 l^4=0\ ,
    \end{align}
    representing a cubic polynomial in $r_h^2$. The discriminant condition then determines that above a critical length scale $l_c$ (for fixed $\Lambda$) or above a critical pressure $P_c$ (for fixed $l$) given by
    \be
    l_c=\frac{\sqrt{\frac{13}{\Lambda}-\frac{5 \sqrt{10}}{\Lambda}}}{3 \sqrt{6}}\ ,\qquad P_c=\frac{5 \sqrt{10}-13}{432 \pi  l^2},
    \ee
    there will be no divergence in the heat capacity, and furthermore all black hole phases become thermodynamically stable. The latter value of $P_c$ can be seen to coincide with the critical pressure \eqref{crit} determined from the equation of state. Note that, contrary to what is claimed in \cite{fan2017}, the coexistence pressure $P=P_{\text{coex}}$ on the $P-V$ diagram constructed from the equal area law
    \be\label{maxwell}
    \int_{v_s}^{v_l}\left(P(v)-P_{\text{coex}}\right)dv=0,
    \ee
    is exactly in correspondence with the pressure at which the small and large black hole phases coexist on the $G-T$ diagram. Therefore, one obtains the same phase diagram from either the $P-V$ curve or the $G-T$ curve, and no redefinition of thermodynamic variables is required. That this is the case is not surprising since the equal area law condition is equivalent to the statement that the two phases have the same free energy, and no new information is contained in the $P-V$ diagram (it is just obtained by a rescaling $\Lambda\rightarrow -8\pi P$). We suspect that the earlier observation that the two coexistence lines differ is due to a numerical error, since in our case demonstrating that \eqref{maxwell} holds requires computing the integration bounds $(v_s,v_l)$ to a precision of at least $10^{-8}$, otherwise the integral does not vanish at the same pressure as the $G-T$ diagram would suggest.
    
    \subsubsection{Critical exponents}
    
    Critical exponents govern the behavior of various thermodynamical parameters near a critical point, and separate physical systems into universality classes under which the scaling behavior of these parameters is identical, even if the underlying microscopic structure of the systems are vastly different. In terms of the reduced temperature $t\equiv T/T_c-1$, they are defined through the scaling behavior of the following quantities:
    \begin{align}
    C_V&\propto |t|^{-\alpha}\ ,\quad \text{g}= v_l-v_s\propto |t|^{\beta}\ ,\\
     \kappa_T=-\dfrac{1}{V}\dfrac{\partial V}{\partial P}\bigg|_T&\propto|t|^{-\gamma}\ ,\qquad |P-P_c|\propto |V-V_c|^{\delta}\ .\nonumber
    \end{align}
    $C_V$ is the heat capacity at constant volume defined by \eqref{cv}, g is the order parameter (here the difference in volume of the large and small black hole phase), $\kappa_T$ is the isothermal compressibility, and $|P-P_c|$ is the behavior of the pressure near the critical point.
    Since the entropy $S=\pi r_h^2$ is independent of temperature, one automatically finds that the critical exponent governing the behavior of the heat capacity near the critical point is $\alpha=0$.\\
    
\begin{figure}[h]
    	\centering
    	\includegraphics[width=0.48\textwidth]{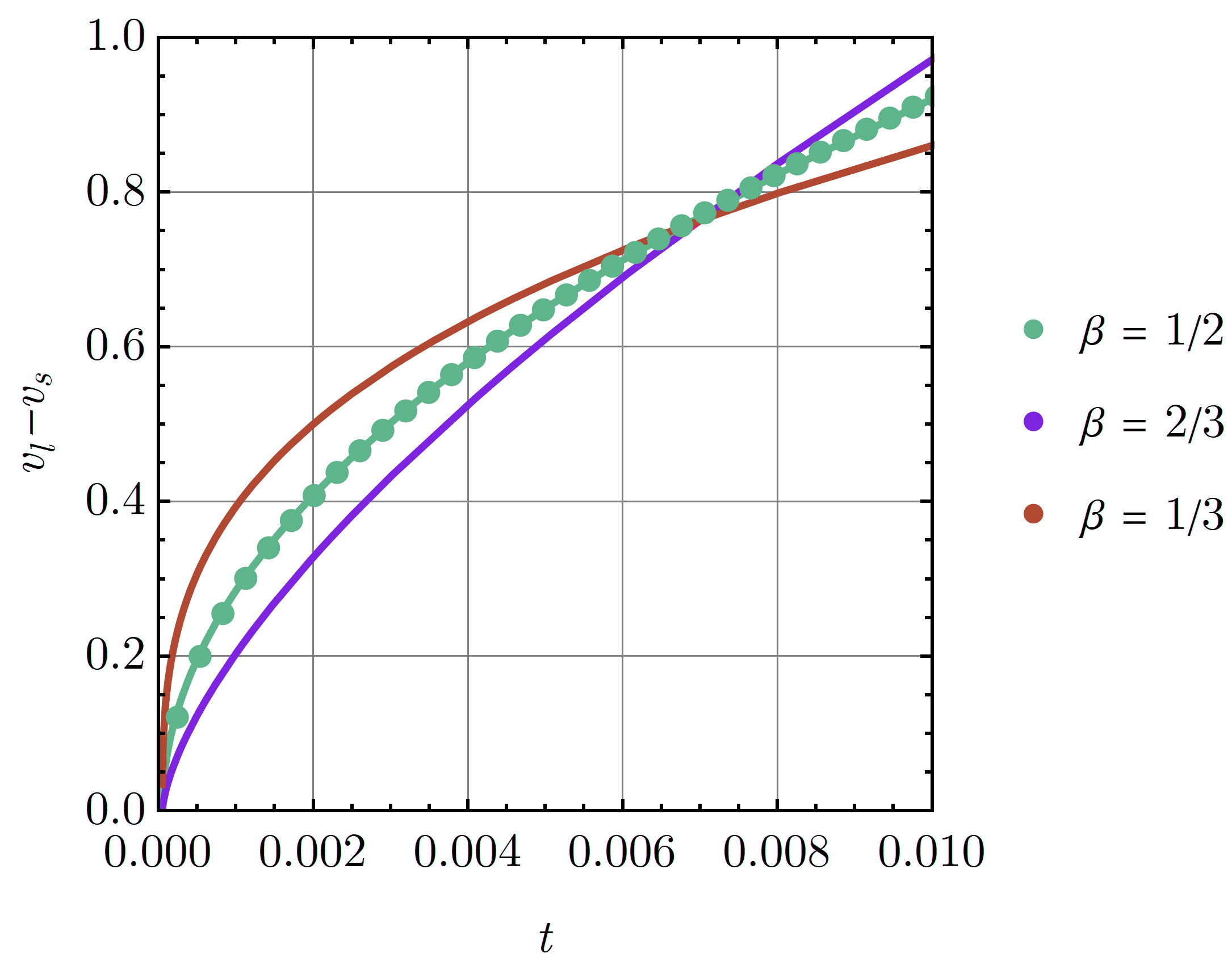}
    	\caption{\small{Scaling behavior of the order parameter $v_l-v_s$ as a function of the reduced temperature $t$ near the critical point. Dots represent numerical data while solid lines represent possible values of the critical exponent $\beta$. }}
    	\label{exponent}
    \end{figure}

    The exponent $\beta$ is computed by examining the difference in size of the small and large black hole phases near the critical point. This cannot be done analytically since the horizon size $r_h$ alone is given by a solution to a 5th degree polynomial, and the functions $G=G(r_h,\Lambda,l)$ and $T=T(r_h,\Lambda,l)$ cannot be solved analytically to find the location of the phase transition. Instead, we proceed numerically and compute the behavior of the order parameter $v_l-v_s$ near the critical point $t=0$. This is done directly from the free energy $G$ and temperature $T$, so that the result is independent of any particular identification of $\Lambda$ and does not require using Maxwell's equal area law to find the coexistence line on the $P(V)$ diagram. The result is shown in Fig.~\ref{exponent} which clearly indicates that $\beta=1/2$. 
   \\
    
    The isothermal compressibility is determined by plotting $\kappa_T$ vs. $t$, which can also be done analytically since
    \be
    \kappa_T\sim-\dfrac{1}{V}\left(\dfrac{\partial V}{\partial r_h}\right)\left(\dfrac{\partial P}{\partial r_h}\right)^{-1}\!\bigg|_T\ .
    \ee 
    It is straightforward to check that $\kappa_T\propto |t|^{-1}$ and therefore $\gamma=1$.
    \\
    
    Finally, the exponent $\delta$ can be determined by evaluating $|P-P_c|$ using \eqref{eq:pressure} and \eqref{crit} and plotting against $|V-V_c|$. We omit the resulting expression as it is lengthy and insightful. This can be done analytically by straightforward substitution, and a parametric plot of $|P-P_c|$ vs. $|V-V_c|$ clearly shows that $\delta=3$.
    \\
    
    We have therefore determined that while the critical ratio $P_c v_c/T_c$ does indeed deviate slightly from the usual value of $3/8$ for the Hayward-AdS black hole (using either the thermodynamic volume $V$ or reduced volume $v$), the critical exponents maintain their mean-field theory values of
    \be\label{exponents}
    \alpha=0\ ,\quad \beta=\frac{1}{2}\ ,\quad\gamma=1\ ,\quad \delta=3\ .
    \ee
    As in other asymptotically AdS examples, this is quite surprising since in this case the equation of state clearly differs from that of the van der Waals fluid, yet the behavior near the critical point exhibits a universal behavior governed by \eqref{exponents}.

\subsection{Minkowski $\Lambda=0$} \label{flat}
	
	Asymptotically flat ($\Lambda=0$) black hole spacetimes admit a straightforward definition of both geometric and thermodynamic variables which enter into the first law, obtained through a Hamilton variation or equivalent covariant phase space formulation. However, the notion of thermodynamic equilibrium is more subtle compared to the asymptotically AdS case. There is no longer an effective potential which naturally confines radiation, and a black hole of any size is generically thermodynamically unstable with a negative specific heat capacity. Since physically reasonable sizes of black holes, which span masses from $10^1$ to $10^{10} M_{\odot}$, have corresponding evaporation timescales on the order of $10^{74}$ to $10^{104}$ seconds, one way to approach the issue is to simply consider the system as being in a state of approximate thermal equilibrium over observationally relevant timescales. This assumption is certainly valid for physical-process interpretations of the first law as applied to physically realistic scenarios  where the back-reaction from both Hawking radiation and infalling matter can reasonably be ignored. However, for holographic applications one can no longer make this approximation since the relevant state space is populated by distinct global configurations labelled by different values of the asymptotic mass, and the physically relevant timescale in the boundary theory may correspond to a bulk timescale for which evaporation cannot be ignored. Therefore, it is useful to introduce another mechanism to define the equilibrium ensemble in asymptotically flat spacetimes.
	\\
	
\begin{figure}[h]
		\centering
		\includegraphics[width=0.485\textwidth]{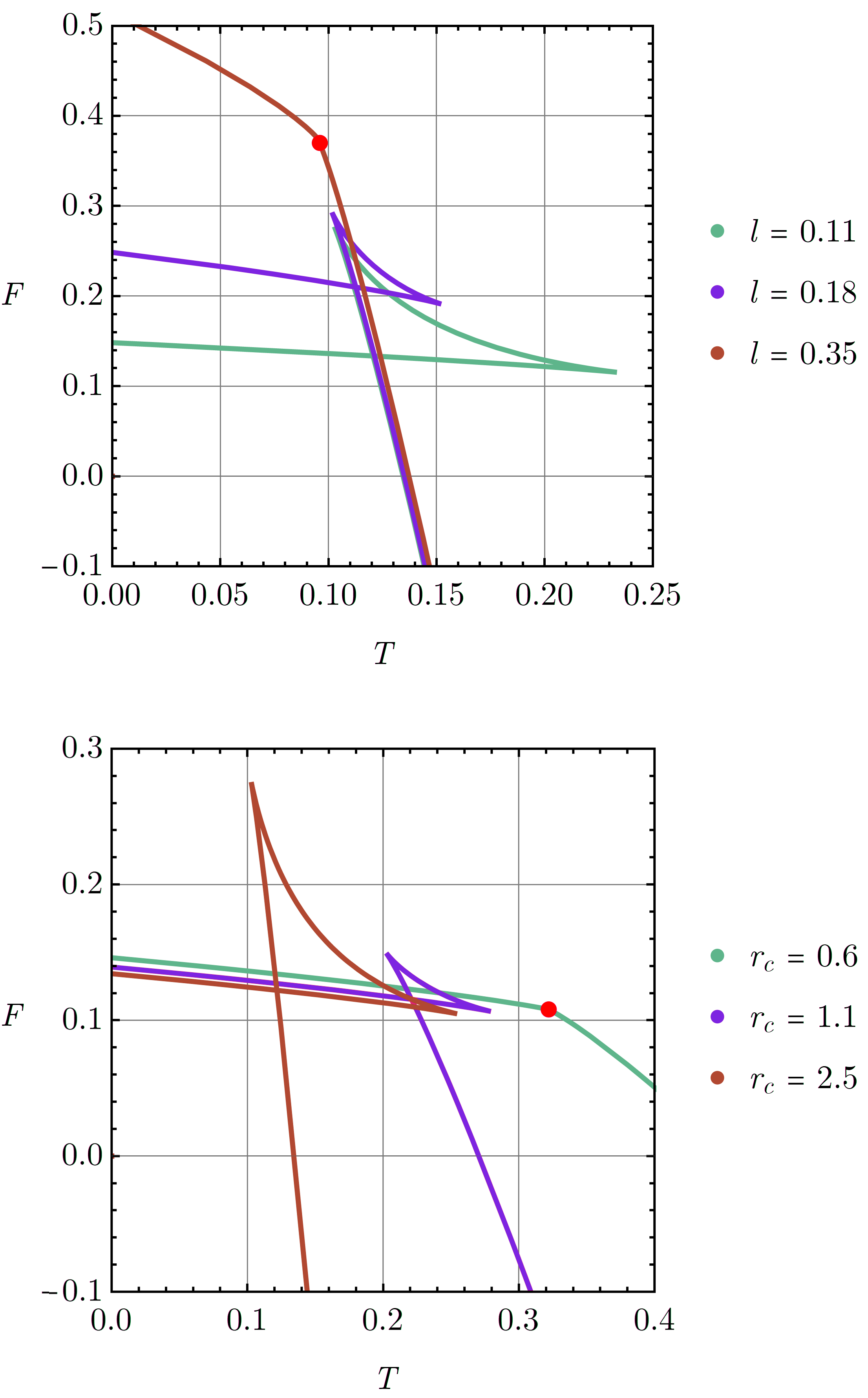}
		\caption{\small{Helmholtz free energy $F$ as a function of temperature $T$ for the asymptotically flat Hayward black hole. {\bf Top:} Fixed $r_c=2$ for various $l$. The onset of a first-order phase transition from a small to large black hole is marked by a red dot at the critical temperature $T_c\sim 0.095$. {\bf Bottom:} Fixed $l=0.1$ for various $r_c$. The critical temperature occurs at $T_c\sim 0.322$.}}
		\label{GT2}
	\end{figure}

	This is most straightforwardly accomplished by the introduction of a ``cavity" representing fixed thermodynamic data on a compact codimension-2 surface outside of the black hole. As described in the previous section, this amounts to introducing a boundary in the Euclidean section where thermodynamic data is specified. Taking the $\Lambda\rightarrow 0$ limit of \eqref{eq:temp-eucl} we recover the equilibrium temperature for the Hayward-Minkowski black hole embedded in an isothermal cavity
	\begin{align}
	\mathcal{T}=\frac{r^6_c(r^3_h-3l^2r_h)}{4\pi\big(r^3_cr^2_h+l^2(r^3_c-r^3_h)^2\big)\sqrt{\frac{r^2_cr^2_h(r_c-r_h)-l^2(r^3_c-r^3_h)}{r^3_cr^2_h-l^2(r^3_c-r^3_h)}}},
	\end{align}
	and the mean thermal energy of the ensemble, by taking the same limit of Eq.\eqref{eq:energy-eucl},
	\be
	E=r_c-r_c \sqrt{\frac{l^2 \left(r_h^3-r_c^3\right)+r_c^2 r_h^2 (r_c-r_h)}{l^2 \left(r_h^3-r_c^3\right)+r_c^3 r_h^2}}.
	\ee
	The free energy can then be computed as $F=\tilde{T} I_E = E-\tilde{T}S$. We find
	\begin{align}
	F&=-\frac{r_c^6 r_h^3 \left(r_h^2-3 l^2\right)}{4 \left(l^2 \left(r_h^3-r_c^3\right)+r_c^3 r_h^2\right)^2 \sqrt{\frac{r_c^2 r_h^3}{l^2 \left(r_c^3-r_h^3\right)-r_c^3 r_h^2}+1}}+\nonumber\\
	&\quad+r_c-r_c \sqrt{\frac{r_c^2 r_h^3}{l^2 \left(r_c^3-r_h^3\right)-r_c^3 r_h^2}+1},
	\end{align}
	which can be plotted parametrically as a function of $r_h$ as before, shown in Fig.~\ref{GT2}. As expected, the behavior is qualitatively similar to the anti-de Sitter case, with a small-large black hole phase transition occurring below a critical value of $l$. Again the $F=0$ phase corresponding to empty Minkowski space is inaccessible due to the finite fixed value of the magnetic charge $Q_m$ required to have $l>0$. Unlike the AdS case there is no notion of thermodynamic pressure for the system, so there is no analogy between the phase structure observed here and mean-field theory systems. The onset of the small-large transition is also controlled now by the cavity radius $r_c$, which when all other parameters are held fixed determines the relative scale between the black hole, cavity, and cosmological horizon.
	
\subsection{de Sitter $\Lambda>0$} \label{desitter}

	We finally turn to the asymptotically de Sitter ($\Lambda>0$) case, which is at the same time the most technically challenging and astrophysically relevant, as there is incontrovertible evidence for both the existence of black hole-like objects \cite{abbott2021} and for the accelerated expansion of the Universe \cite{garnavich1998,goldhaber2009a,planckcollaboration2016}. The latter implies an asymptotically de Sitter geometry as being the relevant background for astrophysical black holes. Therefore, understanding black hole thermodynamics in asymptotically de Sitter spacetimes is important not only due to their observational relevance, but also for applications in the emerging field of de Sitter space holography \cite{strominger2001b,larsen2002,balasubramanian2003}.
	\\
	
	Though the first law can be readily generalized to cases where $\Lambda\neq 0$, dS space presents a host of unique issues which are absent in AdS. The most salient is the presence of a cosmological horizon, which radiates at a much larger temperature than all but the smallest black holes, necessarily placing the system out of equilibrium due to the heat flux between the two horizons. Another issue with de Sitter spaces is a lack of globally timelike Killing vector field with which to associate the mass, making the construction of conserved charges difficult \cite{bousso2002,balasubramanian2002,anninos2011}. The masses which enter into the usual forms of the first law are defined for spacetimes which are asymptotically flat (ADM) or stationary (Komar). In de Sitter, one can recover stationarity by working in the static patch, but then the Killing vector $\xi^a$ that would be used to define the mass becomes spacelike outside the cosmological horizon, rendering the mass conserved in space rather than time \cite{ghezelbash2002c}. The variation of the would-be ADM mass is given by a boundary integral over an $\mathcal{S}^2$ at infinity
	\be
	\delta M=-\dfrac{1}{16 \pi}\int_{\infty}\!\! dS\, n_cB^c[\partial / \partial t]+\cdots\ , \nonumber
	\ee
	but for the region outside the cosmological horizon the Killing vector $\partial/\partial t$ is spacelike, and $M$ cannot support its usual Noether charge interpretation as being the conserved quantity associated with time-translation invariance. Therefore while a variation resembling the first law exists, one cannot straightforwardly interpret the variables entering it as the usual thermodynamical ones. The notion of a vacuum state is also problematic in de Sitter, since the global spacetime is non-stationary \cite{mottola1985,allen1985,goheer2003} and the vacuum state is not even known to be stable \cite{brown1988,anderson2014,firouzjaee2015}.
	\\
	
	Various approaches aside from the Euclidean path integral adopted here have been developed to circumvent these difficulties. One is the {\it effective temperature} approach, where a single temperature $T_{\text{eff}}$ (which depends on both the cosmological and event horizon) is assigned to the entire spacetime \cite{zhangli-chun2010}. This enables one to establish a ``first law" which accounts for the presence of both horizons, but suffers from the fact that $T_{\text{eff}}$ lacks a clear physical interpretation and the system still appears out of equilibrium to a local observer. Another approach is to consider only subsets of the parameter space where the two horizon temperatures are equal \cite{mbarek2019}, in which case equilibrium is trivially established at the price of being limited to a measure-zero subset of possible configurations. One also requires sufficiently many ``charges" to make the temperatures equal, which is not possible for ordinary Schwarzschild-de Sitter black holes.
	\\
	
	\begin{figure}[h]
		\centering
		\includegraphics[width=0.485\textwidth]{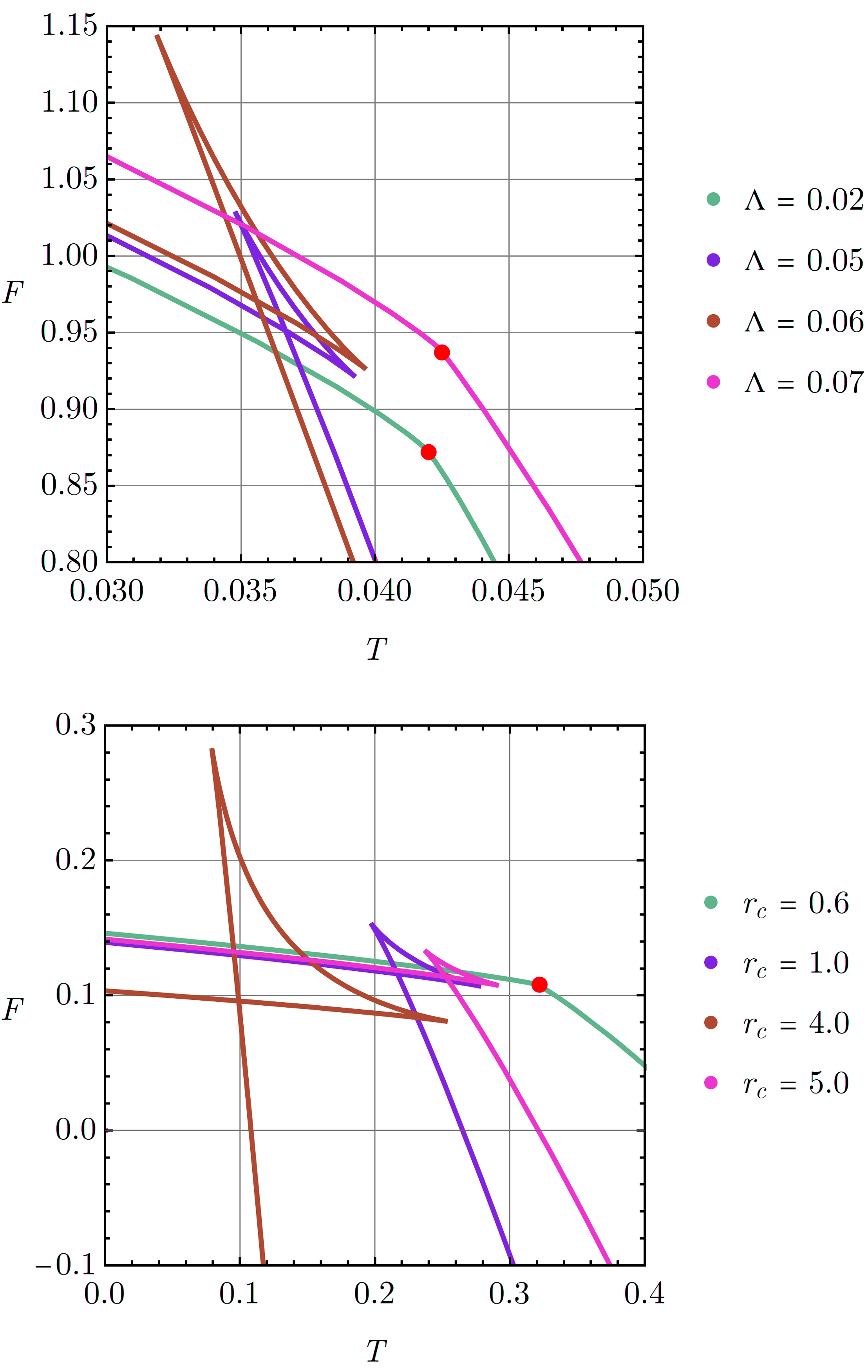}
		\caption{\small{Helmholtz free energy $F$ as a function of temperature $T$ for the Hayward-dS black hole. {\bf Top:} Fixed $l=0.1$ and $r_c=4$ for various $\Lambda$. Two critical points emerge, at $T_{c1}\sim 0.042$ and $T_{c2}\sim0.043$. {\bf Bottom:} Fixed $\Lambda=l=0.1$ for various $r_c$. The onset of a first-order phase transition from a small to large black hole is marked by a red dot at the critical temperature $T_c\sim 0.0315$.}}
		\label{GT3}
	\end{figure}
	
	As in the asymptotically flat case, the Euclidean path integral with finite boundary furnishes an equilibrium ensemble which can be used to study some features of static-patch thermodynamics in de Sitter space. The thermodynamic quantities obtained in Sec.\ref{euclidean} remain valid when $\Lambda>0$, provided that the boundary at $r_c$ is placed between the event and cosmological horizons, such that $r_h<r_c<r_{\text{cosmo}}$ where $r_{\text{cosmo}}$ is the largest real root of the metric function $f(r)$. Without the boundary, there is no choice of periodicity $\beta$ which can eliminate the conical singularity at the event and cosmological horizons simultaneously, except in the degenerate Nariai limit where $r_h=r_\text{cosmo}$. Since $\Lambda>0$ implies that $P<0$, the quantity conjugate to $\Lambda$ should be interpreted as a {\it tension} rather than a pressure. Despite the region of the static patch containing the cosmological horizon being effectively excised from the Euclidean section in this way, the effect of $\Lambda$ still manifests in the required cavity temperature, which observes a significant blueshift when the cavity approaches the cosmological horizon.
	\\

\begin{figure}[h]
		\centering
		\includegraphics[width=0.48\textwidth]{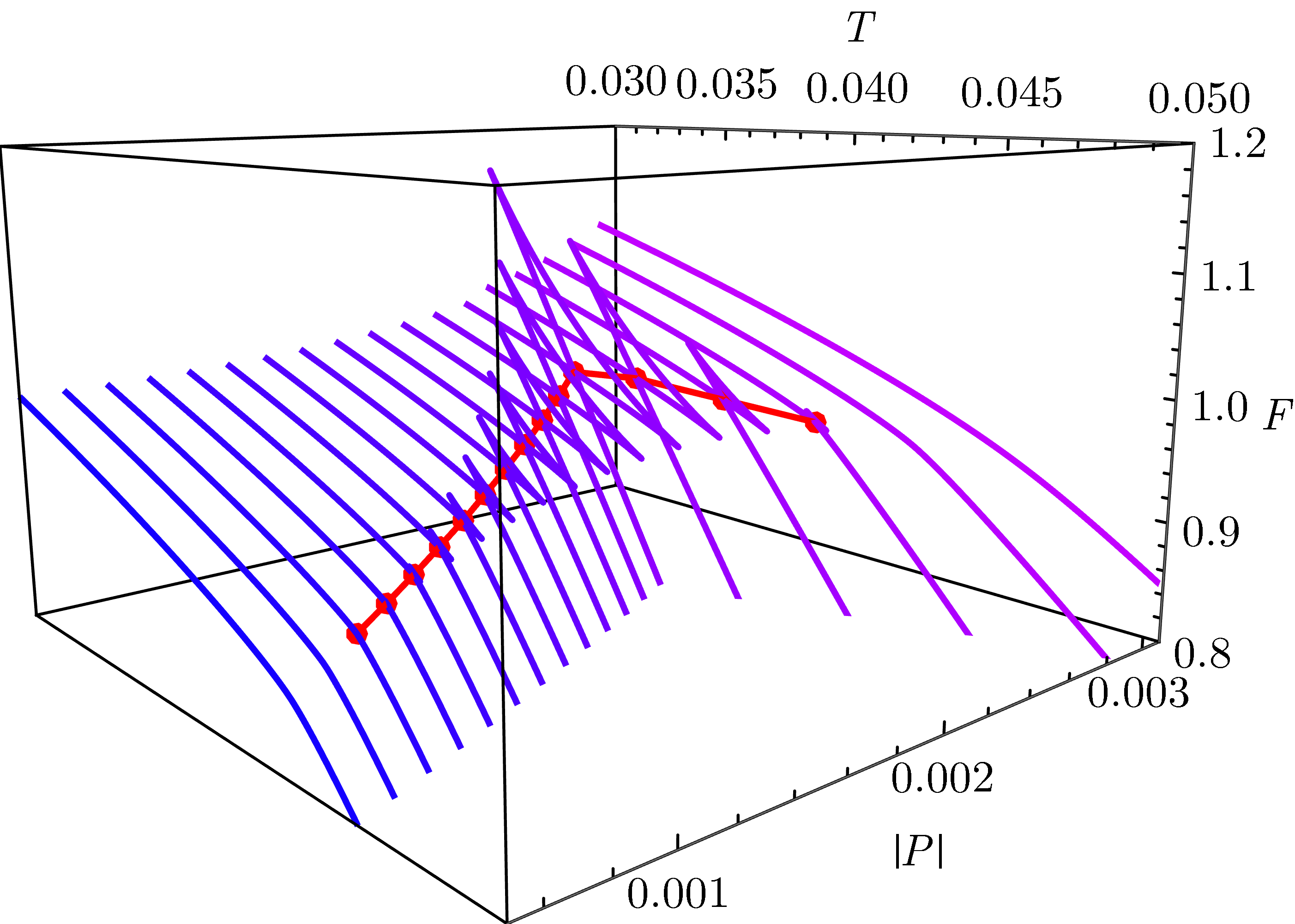}
		\caption{\small{Helmholtz free energy $F$ as a function of temperature $T$ and pressure $P$ for the Hayward-dS black hole, demonstrating the formation of a swallowtube---a compact region in the parameter space where a small-large transition occurs.}}
		\label{FTP1}
	\end{figure}

\begin{figure}[h]
		\centering
		\hspace{10pt}\includegraphics[width=0.45\textwidth]{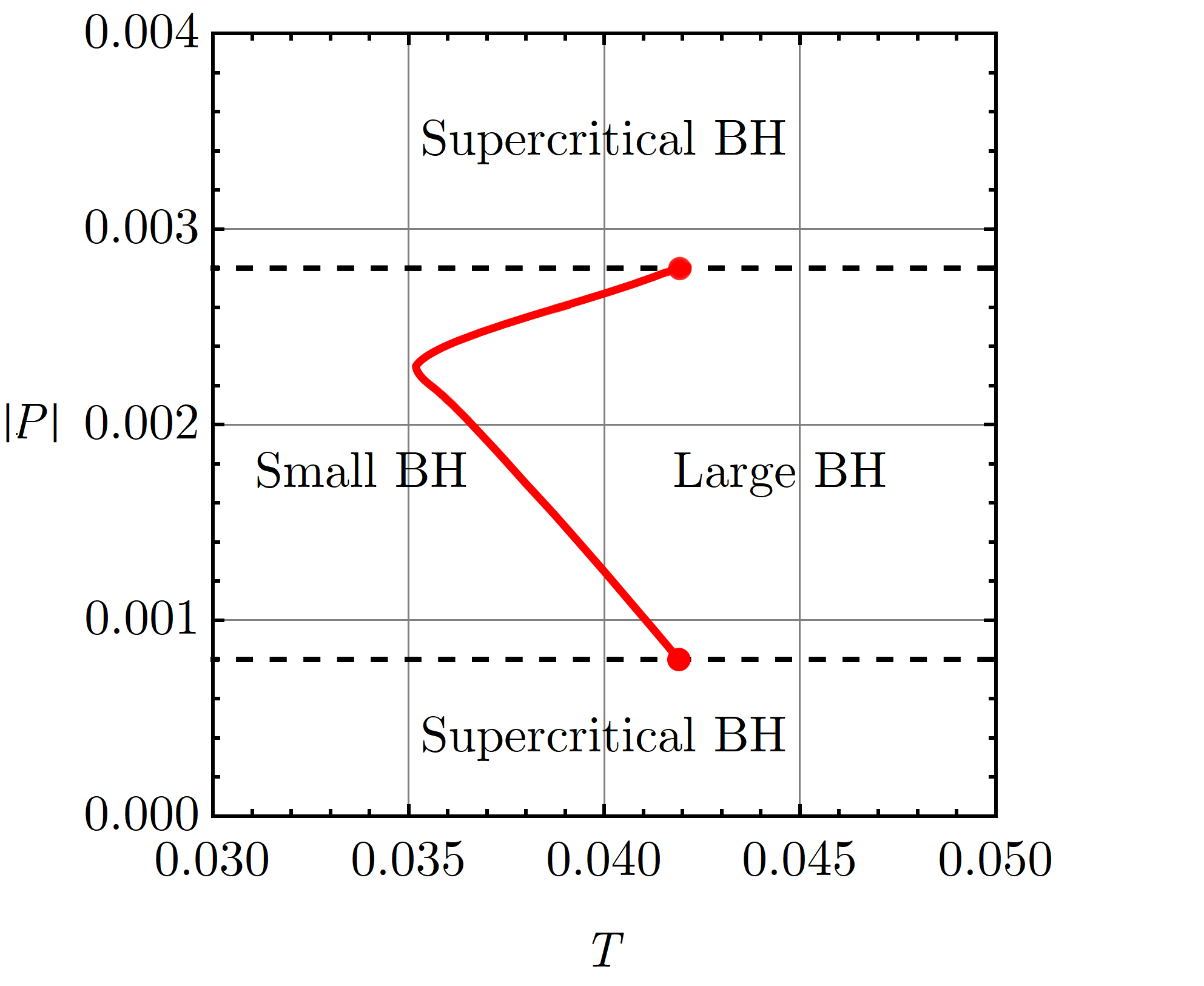}
		\caption{\small{Coexistence line for the Hayward-dS black hole, terminating at two second-order critical points. The system becomes supercritical outside of the region bounded by the dashed black lines.}}
		\label{coex2}
	\end{figure}

	The free energy is obtained from the on-shell Euclidean action as

	\begin{align}
	F&=\frac{l^2 \left(\Lambda r_h^2-3\right)^2+3 \Lambda r_h^4-3 r_h^2}{12r_h}\\
	&\qquad\qquad -\dfrac{4 r_c \left(\mathcal{Y} \sqrt{3-\Lambda r_c^2}+\Lambda r_c^2-3\right)}{12}\nonumber\ ,
	\end{align}
     which is displayed in Fig.~\ref{GT3} as a function of the equilibrium temperature. 
	The presence of a cosmological horizon significantly alters the observed phase structure of the Hayward black hole. As in the AdS case, one can observe a critical point at a maximal pressure ($\Lambda_c\sim 0.07$) below which a line of first-order small-large black hole transitions occurs. As before, above this pressure the system smoothly transitions in size as the temperature increases. However, in the dS case there emerges a second critical point at a lower critical value of $\Lambda\sim 0.02$, below which there is again no phase transition. This is a significant departure from the behavior observed in the asymptotically AdS case, where the small-large transition persists to arbitrarily small $\Lambda$. Fig.~\ref{FTP1} illustrates this more clearly by showing a number of constant-pressure slices. The corresponding coexistence line is shown in Fig.~\ref{coex2}.
	\\

	Unlike AdS black holes in the extended phase space, whose phase structure typically resembles (and is sometimes exactly in correspondence with) the van der Waals fluid, the phase diagram in Fig.~\ref{coex2} more closely resembles that of something akin to the FCC transition of pure solid iron, or quantum critical points in e.g. the transverse field Ising model or a non-Fermi metal \cite{metlitski2015}. However, the phase diagram of Hayward-dS (and other asymptotically dS examples) cannot be in exact correspondence with such materials, because there is no critical point at the cusp of the $P-T$ curve, which is usually at $T=0$ for quantum critical fluids or represents a triple point. Instead, the cusp marks a {\it smooth} transition between a region where the cavity size is on the scale of the black hole horizon and where it is instead on the scale of the cosmological horizon. Furthermore, the coexistence line here terminates at two second-order critical points. It remains to be seen whether such a phase diagram can be understood in a holographic context or whether its novel features are the result of what is effectively a coupling of the black hole system to an external heat bath. We expect that such phase diagrams, like their AdS counterparts, will become increasingly important as a tool for understanding the phase structure of strongly coupled systems as holographic methods in de Sitter space become more refined.

	\section{Conclusions} \label{discussion}
	
	The study of regular black hole solutions in general relativity is an important step toward understanding generic features of quantum gravity, both from a holographic and effective field theory point of view. Regular black hole solutions which can be sourced by matter coupled to Einstein-Hilbert gravity and manage to evade various no-go theorems are few and far between, with the Bardeen and Hayward model being prototypical examples sourced by nonlinear electrodynamics. While previous work which studied the thermodynamic properties of the Hayward solutions focussed on a specific choice of asymptotic metric and thermodynamic formulation/variables, we considered all three types of Hayward black hole (AdS, flat, and dS), discussed which thermodynamic formulation is most appropriate for each, and studied the resulting phase structure of the solutions.
	\\
	
	In the first part of this work, we examined Hayward-AdS black holes with variable cosmological constant. We treated the minimal length scale $l$ as a fundamental thermodynamic parameter, which may arise more generally from the regularization of the Schwarzschild singularity. We demonstrated a consistent version of the first law of black hole mechanics and Smarr relation based on this choice of variable, and studied the extended phase space thermodynamics that results. We found a second-order critical point which marks the formation of a swallowtail in $G-T-P$ space, and a series of first-order small-large black hole phase transitions below the critical temperature/pressure. The critical ratio $P_c v_c/T_c$ was found to deviate from the ``universal" mean-field theory prediction of $3/8$ seen in most other four-dimensional AdS black holes, both when the ordinary reduced volume is used and the volume obtained from an expansion of the equation of state which allows a direct identification with the van der Waals fluid. Contrary to previous investigations into the extended phase structure of Hayward-AdS black holes, we computed the critical exponents and found that they agree with the predicted mean-field theory values.
	\\
	
	We next studied asymptotically flat and de Sitter Hayward black holes. To define an equilibrium thermodynamic (canonical) ensemble, we fixed boundary-value data at a finite radius cavity outside of the black hole horizon (and inside the static-patch cosmological horizon, if it exists). The equilibrium temperature of the ensemble is determined by the choice that leaves the action stationary with respect to variations of the horizon size. Thermodynamic quantities were computed using a semiclassical Euclidean path integral with appropriate boundary terms, and shown to be consistent with the Hamiltonian formulation where the two methods were equally valid. The phase structure in the asymptotically flat case is similar to that of AdS, with a small-large transition occurring which is controlled by the regularization length scale. In asymptotically de Sitter space, two second-order critical points appear which bound a line of first-order small-large transitions. The equation of state deviates significantly from standard van der Waals behavior, and the swallowtail closes at a finite nonzero value of the pressure (in contrast to AdS). The phase diagram contains two supercritical regions, and we conjecture that novel critical behavior may emerge where the two critical points meet.
	\\
	
	A number of generalizations of our work naturally arise. First, it remains to be understood exactly which classes of black hole solutions (and theories) admit a consistent thermodynamic formulation from both the Hamiltonian and Euclidean framework when matter fields are present. While it is known that perfect fluids can be incorporated into the covariant formalism \cite{iyer1997}, this result has not been generalized past symmetry-inheriting matter fields. Furthermore, the linearity of the Smarr relation for nonlinear electrodynamic theories has only been proven for theories that admit a Maxwell weak-field limit, which nonetheless appears to also hold for the NED Lagrangian which generates the Hayward geometry. Another natural extension of our work would be to consider Kerr-Hayward-de Sitter black holes, representing the next logical step in studying thermodynamic features of models which hope to describe astrophysical black holes.

	\section*{Acknowledgments}
	
	F.S. is supported by the ARC Discovery Grant No. DP210101279. I.S. is supported by an International Macquarie University Research Excellence Scholarship. 
	
	\appendix

	\section{ELECTROMAGNETIC POTENTIALS}\label{sec:app:potentials}
	In this appendix we collect various computations of thermodynamical quantities. We begin with the magnetic potential $\Psi(r)$. The only non-vanishing components of the electromagnetic field tensor are $F_{23}=-F_{32}$. In order to compute the magnetic potential one needs to first determine the magnetic field $H_{\mu}$. From Eq. \eqref{eq:B:def}  
	\begin{align}
		\star B_{\mu\nu}=\frac{1}{2}\left(\epsilon_{\mu\nu 23}+\epsilon_{\mu\nu 32}\right)F^{23}=\epsilon_{\mu\nu 23} B^{23}\ .
	\end{align}
	Using the definition of the invariant 4-volume along with Eq. \eqref{eq:B:def} we have that
	\begin{align}
		\star B_{\mu\nu}=\sqrt{-g}\left[\mu\nu 2 3\right]\mathcal{L}'(\mathcal{F})g^{22}g^{33}F_{23}\ .
	\end{align}
	The only non-vanishing components are
	\begin{align}
		\star B_{01}=-\star B_{10}=-\frac{\mathcal{L}'(\mathcal{F})Q_{m}}{r^2}\ ,
	\end{align}
	and using \eqref{eq:H} then gives 
	\begin{align}
		H_{1}=-\star B_{10}\xi^{0}\Rightarrow \frac{\partial \Psi}{\partial r}=-\frac{\mathcal{L}'(\mathcal{F})Q_{m}}{r^2}\ .\label{eq:int}
	\end{align}
	Next one computes \eqref{eq:LF} using
	\begin{align}
		\mathcal{F}=F_{\mu\nu}F^{\mu\nu}=\frac{2Q^2_m}{r^4}\ ,\label{eq:F}
	\end{align}
	since 
	\begin{align}
		F_{\mu\nu}=-Q_{m}\left(\delta^{\theta}_{\mu}\delta^{\phi}_{\nu}-\delta^{\theta}_{\nu}\delta^{\phi}_{\mu}\right)\sin{\theta}\ .
	\end{align}
	Therefore
	\begin{align}
		\mathcal{L}'(\mathcal{F})=\frac{18\sqrt{\alpha \mathcal{F}}}{(1+(\alpha \mathcal{F})^{3/4})^3}\ ,
	\end{align}
	and upon substituting Eq. \eqref{eq:F} we have 
	\begin{align}
		\mathcal{L}'(\mathcal{F})=\frac{18q^2r^7}{(r^3+q^3)^3}\ .\label{eq:L'F}
	\end{align}
	Using now Eq. \eqref{eq:int} along with the relation \eqref{eq:Qm} for the magnetic charge and Eq. \eqref{eq:L'F} we have after integrating with respect to $r$ that
	\begin{align}
		\Psi(r)=\frac{3q^4(2r^3+q^3)}{\sqrt{2\alpha}(r^3+q^3)^3}+\Psi_{0}\ ,
	\end{align}
	where $\Psi_{0}$ is an integration constant which obeys $\Psi_{0}=0$ if  the magnetic potential vanishes at large distances $r\rightarrow\infty$. 
	\\
	
	We now turn to the potential $K_{\alpha}$ conjugate to the parameter $\alpha$. Starting from Eq. \eqref{eq:Ka-int}, the derivative of $\mathcal{L}(\mathcal{F})$ with respect to $\alpha$ is given by 
	\begin{align}
		\frac{\partial \mathcal{L}}{\partial \alpha}=\frac{6\mathcal{F}^2(-2\alpha \mathcal{F}+(\alpha\mathcal{F})^{1/4})}{\left(\alpha \mathcal{F}+(\alpha\mathcal{F})^{1/4}\right)^3}\ ,
	\end{align}
	and then using Eq. \eqref{eq:F} we obtain
	\begin{align}
		\frac{\partial \mathcal{L}}{\partial \alpha}=\frac{6q^6(-2q^3+r^3)}{a^2(r^3+q^3)^3}\ .\label{eq:dLa}
	\end{align} 
	The integration of Eq. \eqref{eq:Ka-int} can then be performed using the above relation \eqref{eq:dLa} and the potential $K_{\alpha}$ given by Eq. \eqref{eq:Ka} is obtained.
	
	\section{EUCLIDEAN ACTION}\label{sec:app:eucl-action}
	In this appendix, we compute the on-shell Euclidean action for the Einstein-Hilbert-NED theory in the semiclassical approximation. The total action will be
	\be
	I_{Total}=I_{EH}+I_{NED}+I_{B}-I_0\ ,
	\ee
	where $I_{EH}$ is the Einstein-Hilbert action with cosmological constant, $I_{NED}$ is the action of the NED theory, $I_{B}$ are appropriate boundary terms for a well-posed variational principle, and $I_0$ is a background subtraction that serves to normalize the action such that $I_{Total}=0$ for the empty (anti-)de Sitter spacetime.
	\\
	
	The Einstein-Hilbert action with cosmological constant in four dimensions is 
     \begin{align}
	    I_{EH}=-\frac{1}{16\pi}\int\! d^4x\sqrt{g}\left(R-2\Lambda\right)\ .
     \end{align} 	
    We adopt a spherically symmetric ansatz for all dominant saddles contributing to the path integral, with Euclidean metrics of the form
    \begin{align}
    	ds^2=f(r)d\tau^2+f(r)^{-1}dr^2+r^2d\Omega_{2}\ ,
    \end{align}
    so that the Ricci scalar is given by 
    \begin{align}
    	R=\frac{2-f''(r)r^2-4rf(r)-2f(r)}{r^2}\ .\label{eq:Ricci}
    \end{align}
 The reduced action is obtained by explicit integration of the action, which can be written as a sum of two terms: 
 \begin{align}
 	I_{EH}=I_{R}+I_{\Lambda}=-\frac{1}{16\pi}\int d^4x\sqrt{g}\,R,\quad I_{\Lambda}=\frac{1}{8\pi}\int d^4x\sqrt{g} \Lambda .
 \end{align} 
 The Ricci part of the Einstein-Hilbert action gives
 	\begin{align}
 	I_{R}=-\frac{1}{16\pi}\int^{\beta_{h}}_{0}\!d\tau\int^{\pi}_{0}\!d\theta\int_{0}^{2\pi}\!d\phi \int^{r_c}_{r_h}r^2\sin{\theta}\,R(r) \ dr\ ,
 	\end{align}
 where the Euclidean section extends from the horizon at $r_h$ to the cavity at $r_c$, and it is assumed in the de Sitter case that $r_c$ is appropriately chosen to lie between the black hole and cosmological horizon, $r_h<r_c<r_{\text{cosmo}}$. The integration over Euclidean time $\tau$ is over one period $\beta_h$, which is chosen to eliminate the conical singularity in the $\tau-r$ plane at the horizon. After integrating over $\tau$,$\theta$ and $\phi$, one obtains
 	\begin{align}
 	 I_{R}=-\frac{\beta_{h}}{4}\int_{r_{h}}^{r_{c}} r^2R(r)\,dr\ .\label{eq:Iricci-int}
 \end{align}
Inserting \eqref{eq:Ricci} into the above then gives
 \begin{align}
 	\int_{r_{h}}^{r_{c}}r^2\,R(r)\ dr=\int_{r_{h}}^{r_{c}}\left[2-f''(r)r^2-4rf'(r)-2f(r)\right]dr\ ,
 \end{align}
 We define 
 \begin{align}
 	\begin{aligned}
 	G_{0}&=\int_{r_{h}}^{r_{c}}f(r)\,dr, \quad G_{1}=\int_{r_{h}}^{r_{c}}rf'(r)\,dr, \\ &\quad\qquad G_{2}=\int_{r_{h}}^{r_{c}}r^2f''(r)\,dr
 	\end{aligned}
 \end{align}
 Computing $G_{1}$ and $G_{2}$ using the fact that $f(r_h)=0$ and integrating by parts reveals that
 \begin{align}
 	G_{2}&=r^2_{c}f'(r_c)-r^2_hf'(r_{h})-2G_{1}\ ,\label{eq:G2} \\
 	G_{1}&=r_{c}f(r_c)- G_{0}\ .\label{eq:G1} 
 \end{align}
 Inserting \eqref{eq:G1} and \eqref{eq:G2} into \eqref{eq:Iricci-int} then gives
 \begin{align}
 	I_{R}=-\frac{\beta_{h}}{4}\left(2(r_{c}-r_{h})-2r_{c}f(r_c)-r^2_{c}f'(r_c)\right)-\frac{\beta_{h}}{4}r^2_{h}f'(r_h)\ ,
 \end{align}
 and upon identifying the periodicity in $\tau$ with the Killing surface gravity
 \begin{align}
 	\beta_{h}^{-1}=\dfrac{\kappa}{2\pi}=\frac{f'(r_{h})}{4\pi}\label{eq:bh}\ ,
 \end{align}
we finally obtain
 \begin{align}
 	I_{R}=-\frac{\beta_{h}}{4}\left[2(r_{c}-r_{h})-2r_{c}f(r_c)-r^2_{c}f'(r_c)\right]-\pi r^2_{h}\ .\label{eq:Iricci-final}
 \end{align}
Next, the cosmological term $I_{\Lambda}$ is straightforwardly evaluated giving
\begin{align}
	I_{\Lambda}&=\frac{1}{8\pi}\int_{0}^{\beta_{h}}\!d\tau  \int_{0}^{2\pi}\!d\phi\  \int_{r_h}^{r_c} \int_{0}^{\pi} r^2\sin{\theta}\ \Lambda\  dr d\theta\nonumber\\
	&=\frac{\beta_h}{6}\Lambda (r^3_c-r^3_h)\ .\label{eq:Ilambda}
\end{align}
We next compute the Gibbons-Hawking-York boundary term \eqref{eq:Ighy}, evaluated at $r=r_c$ :
	\begin{align}
	I_{GHY}=\frac{1}{8\pi}\int_{\partial \mathcal{M}}\!\! \sqrt{k}\,K=\frac{1}{8\pi}\int_{0}^{\beta_{h}}\!d\tau  \int_{0}^{2\pi}\!d\phi\ \int_0^{\pi}\!d\theta \ \sqrt{k}\,K\ \Big|_{r_c}. \label{eq:GHY}
\end{align}
The induced metric $k_{ab}$ on the boundary hypersurface is
\begin{align}
	ds^2=f(r_c)d\tau+r^2_cd\Omega\ ,
\end{align}
and the square root of the determinant of the induced metric is given by 
\begin{align}
	\sqrt{k}=\sqrt{f(r_{c})}\,r^2_{c}\sin{\theta}\ . \label{eq:ind.metric-determ}
\end{align}
The trace $K$ of the extrinsic curvature $K_{ab}$ evaluated at the boundary is 
\begin{align}
	K=-\frac{2\sqrt{f(r_c)}}{r_c}-\frac{f'(r_c)}{2\sqrt{f(r_c)}}\ .\label{eq:extr-curv}
\end{align} 
Combining \eqref{eq:GHY}, \eqref{eq:ind.metric-determ}, and \eqref{eq:extr-curv} we find
\begin{align}
	I_{GHY}=\frac{\beta_{h}}{2}r^2_c K(r_{c})\sqrt{f(r_c)}\ .\label{eq:Ighy-final}
\end{align} 
Next, we compute the NED action given by \eqref{eq:Imatter} with Lagrangian density \eqref{eq:LF}. Using \eqref{eq:F} we can write $\mathcal{L}(\mathcal{F})$ in terms of $\alpha$ and $Q_M$ as 
\begin{align}
	\mathcal{L}(\mathcal{F})=\frac{24\sqrt{2\alpha}\,Q^3_m}{\left(1+\left(\dfrac{2\alpha Q^2_m}{r^4}\right)^{3/4}\right)^2r^6}\ .
\end{align}
Angular integration then reduces the action to
\begin{align}
	I_{NED}=\frac{\beta_{h}}{4}\int_{r_h}^{r_c}r^2\mathcal{L}(\mathcal{F})\,dr\ ,
\end{align}
and the radial integration gives
\begin{align}
	I_{NED}=2\sqrt{2\alpha}\,Q^3_m\beta_{h}\left[\frac{1}{(2\alpha Q^2_m)^{3/4}+r^3_h}-\frac{1}{(2\alpha Q^2_m)^{3/4}+r^3_c}\right]\ .\label{eq:Imatter-final}
\end{align}
The final term we require is the electromagnetic boundary term, given by \cite{MK:21,MK:22}
\begin{align}
	I_{EMB}=-\frac{1}{16\pi}\int_{\partial M}\sqrt{k}\ \left(\frac{\partial \mathcal{L}}{\partial \mathcal{F}}\right)F^{\mu\nu}n_{\nu}A_{\mu}\ ,
\end{align} 
where $n_{\nu}$ is the unit normal vector to the boundary $\partial M$. This term will vanish since we are integrating over a time slice of the spacetime and the only non-vanishing components of the $F_{\mu\nu}$ are $F_{23}=-F_{32}$. Finally, the term $I_{0}$ represents the action of the empty metric (in the absence of the black hole), which serves to normalize the action so that it vanishes when $r_h=0$ i.e. when no black hole is present. Evaluating the total action for the empty metric (Minkowski, de Sitter, or anti-de Sitter), the subtraction term is given by
\begin{align}
	I_{0}=\frac{\beta r_c}{3}(-3+r^2_c\Lambda)\ .
\end{align} 
\section{CAVITY TENSION TERM}\label{sec:app:lambda}
In this section we displayed the tension $\lambda$ conjugate to the cavity area $A_c$, which appears in both the Smarr formula and the first law in the presence of a cavity. We first replace the areal radius $r_c$ in the total reduced action \eqref{eq:red-action} with the area of the cavity, given by $A_{c}=4\pi r^2_c$. The conjugate potential is then given by 
\begin{align}
	\lambda=\frac{1}{\beta}\frac{\partial I_{r}}{\partial A_{c}}\ .
\end{align}
Explicit calculation yields the following, rather long, expression
 \begin{widetext}
	\begin{align}
		\lambda=\frac{1}{48\pi r_c}\bigg(6(1-r^2_c\Lambda)+\frac{2r^2_c\Lambda}{\sqrt{3-r^2_c\Lambda}}\mathcal{X}-2\sqrt{3-r^2_c\Lambda}\mathcal{X}+\frac{r^2_c\sqrt{3-r^2_c\Lambda}\bigg(2l^4(r^3_c-r^3_h)^2\Lambda (-3+r^2_h\Lambda)^2\bigg)}{(3r^2_cr^2_h+\mathcal{Y})^2\mathcal{X}}+\nonumber\\    		    		    		    		+\frac{9r^3_cr^4_h(-3r_h+2r^3_c\Lambda+r^3_h\Lambda)+3l^2r^2_h(4r^6_c\Lambda(-3+r^2_h\Lambda)+2r^4_h(-3+r^2_h\Lambda)^2)+r^3_c(9r_h+r^3_h\Lambda-3r^5_h\Lambda^2)}{(3r^2_cr^2_h+\mathcal{Y})^2\mathcal{X}}\bigg)\ .
	\end{align}
\end{widetext}

	\end{document}